# Natural van der Waals Heterostructural Single Crystals with both Magnetic and Topological Properties

Jiazhen Wu[1], Fucai Liu[2,1*], Masato Sasase[1], Koichiro Ienaga[3], Yukiko Obata[4], Ryu Yukawa[4], Koji Horiba[4], Hiroshi Kumigashira[5,4], Satoshi Okuma[3], Takeshi Inoshita[1,6], and Hideo Hosono[1,7*]

[1] Materials Research Center for Element Strategy, Tokyo Institute of Technology, 4259 Nagatsuta, Midori-ku, Yokohama 226-8503, Japan

[2] School of Optoelectronic Science and Engineering, University of Electronic Science and Technology of China, Chengdu, 610054 China

[3] Department of Physics, Tokyo Institute of Technology, 2-12-1 Ohokayama, Meguro-ku, Tokyo 152-8551 Japan

[4] Photon Factory and Condensed Matter Research Center, Institute of Materials Structure Science, High Energy Accelerator Research Organization (KEK), Tsukuba 305-0801, Japan

[5] Institute of Multidisciplinary Research for Advanced Materials (IMRAM), Tohoku University, Sendai 980-8577, Japan

[6] National Institute for Materials Science, Tsukuba, Ibaraki 305-0044, Japan

[7] Laboratory for Materials and Structures, Institute of Innovative Research, Tokyo Institute of Technology, 4259 Nagatsuta, Midori-ku, Yokohama 226-8503, Japan

*e-mail: fucailiu@uestc.edu.cn, hosono@msl.titech.ac.jp

**Heterostructures possessing both magnetism and topology are promising for the realization of exotic topological quantum states while challenging in synthesis and engineering. Here we report natural magnetic van der Waals heterostructures of $(MnBi_2Te_4)_m(Bi_2Te_3)_n$ that exhibit controllable magnetic properties while maintaining their topological surface states. The interlayer antiferromagnetic exchange coupling is gradually weakened as the separation of magnetic layers increases, and an anomalous Hall effect, that is well coupled with magnetization and shows ferromagnetic hysteresis, was observed below 5 K. The obtained homogenous heterostructure with atomically sharp interface and intrinsic magnetic properties will be an ideal platform for studying the quantum anomalous Hall effect (QAHE), axion insulator states as well as topological magnetoelectric effect.**

**INTRODUCTION**

Magnetic heterostructures have attracted considerable attention in the field of condensed matter physics for novel spintronics and emerging topotronics (*1-15*). Research interest in magnetic heterostructures increased with the emergence of spintronics in 1980s (*1-2*). Well-established deposition techniques for thin film growth, such as molecular beam epitaxy, pulsed laser deposition and sputtering have given rise to the fast growth of this field, where unique properties such as GMR, initially observed in (001)Fe/(001)Cr Superlattices (*4*), and tunneling magnetoresistance (TMR) have been shown to be core technologies for digital information storage (*1*). However, the fabrication of magnetic heterostructures has long been limited by deposition techniques, hindering wide studies of the unique materials systems. Though the transfer method has recently been frequently used to prepare van der Waals heterostructures, it also requires sophisticated techniques with care (*16*).

Recently, heterostructures combined with magnetic layers and topological-insulator layers have led to a rising field for the realization of exotic topological quantum states, including the quantum anomalous Hall effect (QAHE), axion insulator states as well as topological magnetoelectric effect (*5-15*). However, a homogenous heterostructure with atomically sharp interface and intrinsic magnetic properties, believed to be an ideal platform for studying such topological quantum effects, is still experimentally elusive.

In the present work, we report the naturally occurring van der Waals heterostructures of $(MnBi_2Te_4)_m(Bi_2Te_3)_n$, which show controllable magnetic properties and topological surface states (SSs). Single crystals can be prepared using flux method. $MnBi_4Te_7$ (m = n = 1) and $MnBi_6Te_{10}$ (m = 1, n = 2) are identified by x-ray diffraction measurements and the heterogeneous structures are observed directly using scanning transmission electron microscope (STEM). As the interlayer antiferromagnetic (AFM) exchange interactions are gradually weakened with increasing the separation of magnetic layers, the present materials turn into a magnetic order competing system and a ferromagnetic (FM) state could be stabilized below 5 K. As the magnetization has an out-of-plane easy axis, an anomalous Hall effect, that is well coupled with magnetization, is observed. The nontrivial electronic structures of MnBi4Te7, both bulk and surface, are investigated using density functional theory (DFT) calculations, and the present compound is proved to be an AFM topological insulator. The SSs are detected experimentally by ARPES measurements showing a gap (120 meV at 20 K) due to the breakdown of time reversal symmetry. Therefore, the present materials system will provide a platform for investigating various interests in spintronics and topotronics, including spin valve states, QAHE, axion insulator states, two-dimensional van der Waals magnets, etc.

**RESULTS**

MnBi$_2$Te$_4$ was recently reported to be the first intrinsic van der Waals antiferromagnet showing topological nontrivial SS (*10-13*). It shares a similar crystal structure with Bi$_2$Te$_3$, a typical topological insulator (TI). Along the *c* axis, Bi$_2$Te$_3$ is composed of Te-Bi-Te-Bi-Te quintuple atomic layers (QL) held together by van der Waals forces, while MnBi$_2$Te$_4$ is composed of Te-Bi-Te-Mn-Te-Bi-Te septuple atomic layers (SL), where each layer is equal to a QL intercalated by an MnTe layer (Fig. 1 A,B,E,F and Fig. S1 A,B). As the two van der Waals materials have similar lattice constants in the *a-b* plane (the lattice mismatch is only around 1.3%), it would be interesting to check the possibility of synthesizing a natural heterostructure with alternating QL and SL following a new periodicity along the *c* axis (*17*).

Following this assumption, polycrystalline samples of (MnBi$_2$Te$_4$)$_m$(Bi$_2$Te$_3$)$_n$ were prepared using a solid-state reaction route, and their crystal structures are shown in Fig. S1 and Table S1. MnBi$_4$Te$_7$ (m = n = 1) and MnBi$_6$Te$_{10}$ (m = 1, n = 2) are identified as natural van der Waals heterostructures (Fig. S1 F) in which SL are separated by one or two QL, respectively, as spacers (Fig. 1 C,D). As with Bi$_2$Te$_3$ and MnBi$_2$Te$_4$, MnBi$_6$Te$_{10}$ adopts a trigonal structure with a space group of $R\bar{3}m$, while MnBi$_4$Te$_7$ adopts a space group of $P\bar{3}m1$. The direct evidence of heterostructures is obtained by high-angle annular dark field (HAADF)-STEM measurements (for polycrystalline samples), as shown in Fig. 1 E-H and Fig. S2. The atomic resolution images are highly consistent with crystal structures obtained through x-ray diffraction (XRD) measurements and the proposed model (Fig. 1 A-D), clearly showing QL and SL as well as van der Waals gaps. The high degree of crystallinity of the prepared samples can also be confirmed by the selected area electron diffraction (SAED) patterns (Fig. 1 I-L) and the fast Fourier transform patterns of the obtained HAADF-STEM images (Fig. S2). The distinct SAED patterns are mainly the results of different atomic stacking sequences and periodicity along the *c* axis of these compounds.

Single crystals of $MnBi_2Te_4$ and $MnBi_4Te_7$ were grown using a flux-assisted method. These phases were found to be difficult to synthesize because they evolve only at a very narrow temperature range, similar to recent reports on the synthesis of single-crystalline $MnBi_2Te_4$ (*18-21*). The selected pieces of single crystals show shining surfaces and high quality, as indicated by the XRD patterns (Fig. 2). Comparing to $MnBi_2Te_4$, $MnBi_4Te_7$ has more complex diffraction peaks due to its more complex structure containing both QL and SL. The diffraction peak positions and intensities are in agreement with SAED results shown in Fig. 1J, K. The fresh surfaces of the samples were checked by Auger electron spectroscopy and X-ray photoelectron spectroscopy under high vacuum (around $10^{-10}$ Torr), and the results show that the samples are very clean and contain all the proposed elements, *i.e.*, Mn, Bi and Te (Fig. S3). The content of Mn, which is the magnetic center in $MnBi_4Te_7$ is about half of the content in $MnBi_2Te_4$ (Fig. S3D). This is consistent with the results that the former contains half QL and half SL while the latter contains SL only.

It is noteworthy that the preparation of $(MnBi_2Te_4)_m(Bi_2Te_3)_n$ with larger values of m (>2) and n (n>1) would be rather challenging. The degree of difficulty for synthesis increases greatly from $MnBi_2Te_4$ to $MnBi_4Te_7$ and to $MnBi_6Te_{10}$, and we observed only a trace amount of m=3, n=1 phase in our solid-state reaction route for polycrystalline samples. For single-crystal growth, $MnBi_2Te_4$ evolves only at a narrow temperature interval (around 10 °C from 590 °C to 600 °C) (*18-20*), $MnBi_4Te_7$ evolves only within 2 °C near 588 °C, and the growth temperature for $MnBi_6Te_{10}$ is even narrower. Therefore, precise temperature controlling would be the most difficult point for growing the compounds with greater m and n values. Alternatively, based on the present findings, deposition or transfer method might be feasible to prepare $(MnBi_2Te_4)_m(Bi_2Te_3)_n$ with any combination of m and n, though it is also challenging in skills.

The structural modifications are strongly indicative of modulations of the magnetic properties. We first measured the magnetization of polycrystalline $MnBi_2Te_4$ and a mixture of $MnBi_4Te_7$ and $MnBi_6Te_{10}$. The magnetization of $MnBi_2Te_4$ is consistent with a previous report (*18*), showing an AFM transition at $T_N = 25$ K and a spin-flop transition at 3.54 T (Fig. S4 A-B). For the mixture of $MnBi_4Te_7$ and $MnBi_6Te_{10}$, magnetic susceptibility demonstrates an FM transition at $T_C = 12.1$ K for fields greater than 0.1 T (Fig. S4 C-H). However, for fields that are less than 0.1 T, there are two magnetic transitions: an AFM transition ($T_{N1} \cong 13$ K) contributed from $MnBi_4Te_7$ and a spin glass-like transition ($T_{N2} \cong 6.5$ K) probably contributed from $MnBi_6Te_{10}$ or amorphous phases (inhomogeneous QL and SL heterostructures). This field-dependent magnetization displays a saturation value at high fields and magnetic hysteresis at low fields below 5 K, indicating the arising of FM states.

To make the magnetic structures clear, we conducted magnetization measurements on single-crystalline MnBi$_2$Te$_4$ and MnBi$_4$Te$_7$. The two compounds show remarkably different magnetic structures (Fig. 3). For MnBi$_2$Te$_4$, the results are consistent with those for the polycrystalline sample, additionally demonstrating that the magnetization-easy axis is along the *c* direction (*10, 11, 19, 20*). While the interlayer exchange coupling (IEC, which favors AFM states) of MnBi$_2$Te$_4$ is at an intermediate strength, the IEC of MnBi$_4$Te$_7$ becomes much weaker. For MnBi$_4$Te$_7$, at a field of 1 T, the AFM transition disappears and an FM transition occurs at $T_C = 12.5$ K (Fig. 3D). The magnetization is saturated above 1 T for fields either along the *c* axis or the *a-b* plane (Fig. 3 E,F), indicating that the spin orientation can easily be realigned by an external field irrespective of the crystal orientation. However, similar to MnBi$_2$Te$_4$, MnBi$_4$Te$_7$ also exhibits a magnetization-easy axis along the *c* direction (Fig. S5A).

It is noteworthy that the AFM-IEC can still play an important role at small fields as shown in Fig. 3G that an AFM transition takes place at $T_N = 12.7$ K. However, AFM-IEC seems not to be the only interaction that determines the magnetic order, because an obvious upturn was observed below $T_N$, especially for fields applied along the *a-b* plane. The first derivative of magnetic susceptibility over temperature also suggests that there might be another magnetic transition near 2 K (Fig. S6 A-F). The field-dependent magnetization when a field is applied along the *a-b* plane shows a nonlinear curve below 5 K and a small hysteresis at 2 K (Fig. 3H), suggesting an additional magnetic interaction (AMI) that tends to cant the spins away from their out-of-plane direction and favors an FM state. When the field is applied along the *z* axis, much higher hysteresis was observed at 2 K with spin-flip transitions connecting FM phase at high fields and AFM phase at the critical field ($H_c$). Interestingly, the FM states can be maintained even at 0 T, indicating again the existence of AMI. However, the AMI decays rapidly with increasing thermal energy and the magnetic hysteresis disappears above 5 K. Fig. S6 provides a detailed analysis of the magnetic structure. The AFM transitions in MnBi$_2$Te$_4$ and MnBi$_4$Te$_7$ were also observed in their low-temperature heat capacities (Fig. S5B and Fig. S7A). It is important to note that due to the weak IEC, the magnetic properties of MnBi$_4$Te$_7$ can be modified by controlling the applied magnetic field (strength and direction). Also, the IEC can be modified by controlling the spacers (QL), and it gradually decreases from MnBi$_2$Te$_4$ to MnBi$_4$Te$_7$ and to MnBi$_6$Te$_{10}$. In view of these features, we regard the magnetic properties of (MnBi$_2$Te$_4$)$_m$(Bi$_2$Te$_3$)$_n$ as "tunable".

To gain insight into the electronic structure and topology of MnBi$_4$Te$_7$, we carried out first-principles DFT calculations based on the hybrid functional (HSE06) method, which is used widely for the study of small band gap materials. The obtained band structures, with and without spin-orbit coupling (SOC), of bulk MnBi$_4$Te$_7$ are shown in Fig. 4A and Fig. 4B, respectively. When SOC is turned off, Te-5$p$ states (valence bands) and Bi-6$p$ states (conduction bands) are separated by a direct gap of 633 meV at $\Gamma$ point (Fig. 4A), consistent with their balanced charge states, [Bi$^{3+}$]$_2$[Te$^{2-}$]$_3$. However, this simple chemical picture breaks down when SOC is taken into account (Fig. 4B) with Te-5$p$ bands and Bi-6$p$ bands hybridizing near the Fermi level to cause band inversion (band gap = 247 meV). Such band inversion induced by strong SOC suggests that the material is topologically nontrivial (*22*). According to Mong, Essin and Moore, AFM insulators breaking both time-reversal ($\Theta$) and a primitive lattice translation symmetry ($T_{1/2}$) but preserving their combination $S = \Theta T_{1/2}$ (MnBi$_4$Te$_7$ is one such material) can be classified by $\mathbb{Z}_2$ topology (*23*). We carried out a Wilson loop calculation (*24, 25*) for MnBi$_4$Te$_7$ and confirmed that it is indeed an AFM TI with $\mathbb{Z}_2 = 1$.

TIs are known to have SSs associated with their bulk topology (*22*). Figures 4 C and D show the electronic structures of MnBi$_4$Te$_7$ slabs with two surface terminations, QL-terminated and SL-terminated, respectively. Distinct from the gapless SS in Bi$_2$Te$_3$, the SSs of MnBi$_4$Te$_7$ are gapped with a gap of 11 meV (indirect) and 62 meV (direct) for the QL-terminated and SL-terminated surface, respectively. The nonvanishing gap is consistent with the broken time-reversal symmetry, reflecting presumably the interaction between the SSs and the FM layer near the surface, similar to the gapped SS of MnBi$_2$Te$_4$ (another AFM TI) (*10, 19*).

The SS of MnBi$_4$Te$_7$ were measured using angle-resolved photoemission spectroscopy (ARPES) at 20 K and 300 K with an excitation photon energy of 48 eV (Fig. 5), which had been employed for the experimental confirmation of the SS of Bi$_2$Te$_3$, a prototypical three-dimensional TI (*26*). Closely resembling the SS of Bi$_2$Te$_3$ while having a finite energy gap, the observed bands were confirmed to be SS, for which the $\bar{\Gamma}$ point was confirmed through measurements of the Fermi surface in the $k_x$-$k_y$ plane (Fig. S3 G,H). Compared with the calculation results (Fig. 4 C,D), the measured SS (Fig. 5A) were found to be contributed mainly from the SL-terminated surface as indicated by the bump (or cusp) at the valence band maximum. However, at the present stage, we cannot exclude the contributions from the QL-terminated surface, as it should be equally exposed on the mechanically cleaved surface, (there are steps connecting QL and SL). Considering that the QL/SL surface domain size may be much smaller than the photon beam spot size in the present measurements, we speculate that both surfaces were measured. There are several possible reasons that the observed SS look more like the contributions from SL, such as the self-protecting behavior of the topological states as proposed in a previous report (*27*) or a significant larger cross section of SL-SS than that of QL-SS. According to Fig. 4 C,D, the contribution from the SL-terminated surface is much more significant than that from the QL-terminated surface at the Γ point near the gap. This may be another reason that we cannot resolve the two different types of surfaces and the observed SS at the Γ point near the gap is mostly from SL. Given that the photon beam is only focused on QL, its SS would be largely enhanced and observable.

$MnBi_4Te_7$ exhibits surface band gaps of $120 \pm 10$ meV and $90 \pm 10$ meV at 20 K and 300 K, respectively (Fig. 5 B,D), similar to the reported data for $MnBi_2Te_4$ (*10, 19*). The opening of the surface band gap above $T_N$ (12.7 K) indicates that the time-reversal invariance is broken even in the paramagnetic (PM) phase for $MnBi_4Te_7$. Similar phenomena have been frequently observed in intrinsic magnetic topological insulators (*10, 19*) and magnetically doped topological insulators (*28-31*), but no consensus has been reached on the mechanism. Rader and his coworkers proposed that impurity-induced resonance states may destroy the Dirac point and explain the nonmagnetic gap (*28*), while this mechanism was excluded as the primary factor in $MnBi_2Te_4$ (a system similar to ours) (*10*). Mao *et al.* ascribed the PM gap to strong FM-type spin fluctuations (*19*). Actually, we have also observed FM spin fluctuations in $MnBi_4Te_7$ above $T_N$ as indicated by the positive Curie-Weiss temperature (Fig. S6 B,E), therefore the present results may also suggest spin-fluctuations as a possible reason for the SS gap. However, this is an open question requiring further investigations. A theory for the SS when the material is in the PM phase would be very helpful. It should be noted the SSs of $MnBi_4Te_7$ are more complex than those of $MnBi_2Te_4$, as indicated by a flattening of the conduction band and a cusp in the valence band near the $\bar{\Gamma}$ point (Fig. 5A), reminiscent of the more sophisticated SSs of $PbBi_4Te_7$ than that of $PbBi_2Te_4$ (*27, 32*). With gapped nontrivial SSs and tunable magnetic properties, magnetic heterostructures of $(MnBi_2Te_4)_m(Bi_2Te_3)_n$ will be an ideal platform for exploring tunable quantized magnetoelectric phenomena such as QAHE and axion insulator states.

The electrical transport properties of $MnBi_4Te_7$ single crystal are shown in Fig. 6, and they are notably different from those of $MnBi_2Te_4$ (Fig. S7). The compound has a metallic conductivity (Fig. S5C), and the Hall effect shows that the carrier is n-type with a carrier concentration ($n$) of $2.85 \times 10^{20}$ cm$^{-3}$ at 2 K (Fig. S5D), consistent with the ARPES results that Fermi level is not located in the SS Zeeman gap but in the conduction band (Fig. 5). The Hall resistivity has a linear field-dependence at high fields (Fig. 6A), suggesting a single carrier in the present compound. The derived anomalous Hall (AH) resistivity (Fig. 6A) is well coupled with the magnetization curves (Fig. 3I) and shows hysteresis at low temperatures, which is strongly indicative of the possibility of observing QAHE if the Fermi level is tuned in the SS Zeeman gap (*6, 7, 14*). The AH conductivity, which can be expressed as $\sigma_{xy}^A = \rho_{yx}^A/\rho_{xx}^2$ (*33*), is shown to be virtually temperature-independent below 2 K while its temperature-dependence above 2 K follows the same trend as magnetic susceptibility (Fig. 6 B,C). Here $\rho_{yx}^A$ and $\rho_{xx}$ represent the AH resistivity and longitudinal resistivity, respectively. Generally there are three main mechanisms accounting for the AH effect, intrinsic Berry's phase curvature in the momentum space, side jump and skew scattering (*33*). In the present study, the temperature-independent $\sigma_{xy}^A$ and the scaling factor $S_\text{H}$ ($\sigma_{xy}^A = S_\text{H}M$, $M$ is the magnetization, Fig. 6C), and $\sigma_{xx}$-independent $\sigma_{xy}^A$ (Fig. S5F) exclude skew scattering as the primary reason and suggest that the intrinsic Berry's phase curvature is the major contribution to the AH effect (*33-36*). More studies are necessary for elucidation of the details. It should be noted that a small nonlinear coupling of $\rho_{yx}^A$ with $M$ was observed below 5 K near the spin-flip transition (Fig. S8), and the discrepancy ($\Delta\rho_H^A$) is $n$-independent and probably due to the intrinsic contribution from the net Berry's curvature of a noncollinear spin texture (*19, 37*).

The spin-flip transitions can also be reflected in magnetoresistance ($\rho_{xx}$), as shown in Fig. 6 D,E and Fig. S9. At higher temperatures (5–12 K), there are two peaks in $\rho_{xx}$ at $H_{c1}$ and $H_{c2}$, which are critical fields for spin-flip transitions with $H_{c2} = -H_{c1}$ (note that $H_{c1}$ and $H_{c2}$ are essentially the same and their signs only depend on the definition of the sign of the applied magnetic field). The sample is in AFM phase at a field between $H_{c1}$ and $H_{c2}$, and in FM phase at higher fields (Fig. 6F). At lower temperatures (50 mK–2 K), $H_{c1}$ and $H_{c2}$ are on the same side (corresponding to the magnetic hysteresis of $\rho_{yx}^A$), either positive or negative, depending on the sweep direction of the field (Fig. 6F and Fig. S5I), indicating that the sample turns into a spin-valve-like state (*1-3*). $H_{c1}$ and $H_{c2}$ are nearly merged from 0.35 K to 2 K (Fig. 6F), demonstrating that AFM states can survive only within a very narrow range of fields near the critical field. Interestingly, there are plateaus (high-resistance state) in $\rho_{xx}$ between $H_{c1}$ and $H_{c2}$ at temperatures lower than 0.3 K (Fig. 6E), indicating that electrons suffer a higher scattering rate than they do at a lower or higher field. This implies that, different from the FM states at high fields (larger than $H_{c2}$), the material is now in an AFM state, similar to the phenomenon frequently observed for spin valves (*1-3*). However, the magnetoresistance plateaus cannot survive at higher temperatures ($\geq 0.35$ K), where thermal activation may destroy the AFM ordering quickly and the system enters a FM state. Although FM state at a field between $H_{c1}$ and $H_{c2}$ should be more stable (in a lower energy level), it seems that the transition from AFM to FM cannot be automatically happened (below 0.3 K) without sufficient thermal energy to overcome an energy barrier. This is supported by Fig. S9E, where, thermal activation for the transition from the high-resistance state (AFM) to low-resistance state (FM) can be clearly observed (the resistance from 70 mK to 1 K at 0.15 T). It is noted that the plateaus are also shown in AH conductivity (Fig. 6B), resembling axion insulating states (*8, 12, 14*). Therefore, the present system is also a potential platform for creating axion insulators if the Fermi level is tuned into the SS Zeeman gap. In addition, when current flows across the magnetic and nonmagnetic layers, the magnetoresistance effect seems much stronger (Fig. S5H), similar to materials which show giant magnetoresistance (*1, 4*).

**DISCUSSION**

The field- and temperature-dependent magnetic structures of MnBi$_4$Te$_7$ are summarized in Fig. 6F and Fig. S5I. The complex magnetic interactions indicate that MnBi$_4$Te$_7$ can be regarded as a magnetic order competing system. We can express the Hamiltonian regarding to spin structures in an applied magnetic field as, $\mathcal{H} = \mathcal{H}_{\text{IEC}} + \mathcal{H}_{\text{AMI}} + \mathcal{H}_{\text{Z}}(B)$, where $\mathcal{H}_{\text{Z}}(B)$ is the Zeeman energy that depends on the field, *B*. As both $\mathcal{H}_{\text{IEC}}$ and $\mathcal{H}_{\text{AMI}}$ are small in the present system, only a small critical field is required to achieve a sufficient level of $\mathcal{H}_{\text{Z}}(B)$ to align the spins. The competition is mainly between $\mathcal{H}_{\text{IEC}}$ terms that favor the AFM states and $\mathcal{H}_{\text{AMI}}$ terms that favor the FM states, and the AFM states are found to be metastable (Fig. S9E). Both $\mathcal{H}_{\text{IEC}}$ and $\mathcal{H}_{\text{AMI}}$ are weakened with increasing temperature, however $\mathcal{H}_{\text{AMI}}$ decays much faster than $\mathcal{H}_{\text{IEC}}$. The energy difference obtained by DFT calculations between the FM and AFM states is only $\mathcal{E}(\text{FM}) - \mathcal{E}(\text{AFM}) = 0.44$ meV per Mn ion, comparable to a thermal energy of 5 K (Supplementary Method). This small energy difference strongly indicates the possibility of competing FM and AFM states in the present material. There are several candidates for AMI: Ruderman–Kittel–Kasuya–Yosida (RKKY) interaction mediated by conduction carriers (*38*), Dzyaloshinskii-Moriya interaction due to the breakdown of inversion symmetry (*39*), and van Vleck mechanism-mediated FM interaction, as reported in diluted magnetic doped TI (*40*). However, further investigation is needed to determine the exact mechanism. It is noted that this competing situation is not found in MnBi$_2$Te$_4$ (Fig. S7E), in which the AFM-IMC is much stronger and may exceed the AMI. Therefore, the competing magnetic order in MnBi$_4$Te$_7$ may induce unexplored topological states.

In summary, the exotic magnetic structures of the present materials will not only lead to fundamental interests in magnetism, but also provide a new platform (by combining the non-trivial SSs) for topotronics toward quantized magnetoelectric phenomena. Additionally, successful isolation of the van der Waals materials will provide brand new opportunities to study the interplay between magnetism and topology down to two-dimensional limits (*41*).

**MATERIALS AND METHODS**

**Crystal growth.** Polycrystalline samples of MnBi$_2$Te$_4$ and a mixture of MnBi$_4$Te$_7$ and MnBi$_6$Te$_{10}$ were synthesized using a solid-state reaction route. Stoichiometric amounts of the elements (Mn:Bi:Te = 1:4:7 for the mixture) were loaded inside a carbon-coated silica tube and sealed under vacuum and then pretreated using a hydrogen flame. The fused ingot was crushed, ground into a powder and pelleted inside an Ar-filled glove box. Then, the pellet was sealed in vacuum inside a carbon-coated silica tube and annealed at 550 °C for 5 days and finally water-quenched. For the growth of $MnBi_2Te_4$ and $MnBi_4Te_7$ single crystals, a similar procedure was performed before heating in a furnace. The starting element composition ratio was Mn:Bi:Te = 1:6:10 in both cases. Further reduction of Mn content also worked well. The key point for the synthesis is controlling the growth temperature, which was reported to be as narrow as 10 °C (*17-20*). The present synthesis was conducted by heating up to 1000 °C and maintaining it for 3 h, and cooling down to 600 °C for 10 hours, and then cooling to $T_{end}$ over one week, and finally quenching using ice-cold water. $T_{end}$ should be above the melting point of $Bi_2Te_3$. It was set to 590 °C for the growth of $MnBi_2Te_4$ and 588 °C for $MnBi_4Te_7$. The growth method can be regarded as a flux ($Bi_2Te_3$) method (*20*). Single crystals with laminated structures were peeled and cleaved from the grown bulks. As the quenching temperature is higher than the melting point of $Bi_2Te_3$, no $Bi_2Te_3$ single-crystal pieces could be found. The $Bi_2Te_3$ obtained after quenching shows bad crystallinity, as indicated by broad and asymmetric XRD peaks.

**Sample characterization.** Powder XRD measurements were performed using an X-ray powder diffractometer with Cu K$\alpha$ radiation (Bruker D8 Advance), and the results were refined using the GSAS package (*42*). Auger electron spectra (AES) were obtained with 10 keV primary electrons using a Scanning Auger Nanoprobe System (PHI 710; ULVAC-PHI). High-angle annular dark-field scanning transmission electron microscopy (HAADF-STEM) images were obtained using a JEOL JEM-ARM200F atomic resolution analytical electron microscope operated at an accelerating voltage of 200 kV. To observe the lattice periodicity along *c* direction, TEM specimens were prepared using the ion milling technique. Firstly, the powder was mixed with epoxy resin and loaded into a stainless pipe with a diameter of 3 mm. Then the pipe was thinned by mechanical polishing and dimpling until the central part reached a thickness of 40 μm. Finally, it was further thinned by ion milling using 4 keV Ar ions.

**Physical properties measurements.** Magnetization was measured using a Magnetic Property Measurement System (MPMS SQUID VSM Quantum Design). Electrical transport properties (resistance, magnetoresistance and Hall resistance) and heat capacity measurements above 2 K were performed using a Physical Properties Measurement System (PPMS, Quantum Design). The electrical transport properties below 1 K were measured using a $^3$He-$^4$He dilution refrigerator. For the measurements of the electrical transport properties, the magnetic field was applied along the *c* axis of the van der Waals crystals.

**Surface band structure characterization.** ARPES measurements were carried out using horizontally polarized synchrotron radiation light at the BL–2A MUSASHI beam line of the Photon Factory, KEK. Samples were cleaved *in situ* under an ultrahigh vacuum of $1\times10^{-10}$ Torr, and the experimental data were collected using a Scienta SES-2002 electron energy analyzer. The sample temperature during the ARPES measurements was set to 20 K and 300 K, with a total energy resolution of 30 meV at a photon energy ($h\nu$) of 48 eV. The size of the photon beam is around 300 μm in diameter.

**Theoretical calculations.** Ab initio electronic structure calculations were carried out using the projector augmented wave (PAW) method with a plane wave basis, as implemented in the Vienna ab initio simulation package (VASP) (*43, 44*). The exchange-correlation potential was treated in the screened Heyd–Scuseria–Ernzerhof hybrid functional scheme in the HSE06 parameterization (*45, 46*). The plane waves were cut off at 350 eV, while the k-sampling mesh was chosen to be 6×6×2 and 5×5×1 for bulk and slabs, respectively. The Wilson loop calculation to determine the topological index was carried out using a tight-binding Hamiltonian in the Wannier function basis derived from the ab initio calculation (*24, 25*). The ground state of the system was assumed to be AFM.

**SUPPLEMENTARY MATERIALS**

Section S1. Data analysis methods for electrical transport measurements.
Fig. S1. Crystal structures and XRD patterns.
Fig. S2. Additional STEM data.
Fig. S3. Additional surface characterization.
Fig. S4. Magnetization of polycrystalline samples.
Fig. S5. Additional physical properties of MnBi$_4$Te$_7$ single crystal.
Fig. S6. Additional information for the magnetization of MnBi$_4$Te$_7$ single crystal.
Fig. S7. Additional physical properties of MnBi$_2$Te$_4$ single crystal.
Fig. S8. Magnetization dependence of $\rho_{yx}^A$ for MnBi$_4$Te$_7$ single crystal.
Fig. S9. Additional electrical transport properties below 1K for MnBi$_4$Te$_7$ single crystal.
Table S1. Crystal structure parameters of Mn-Bi-Te compounds.

**REFERENCES AND NOTES**

1. I. Žutić, J. Fabian, S. D. Sarma, Spintronics: Fundamentals and applications. *Rev. Mod. Phys.* **76,** 323-410 (2004).
2. A. Brataas, A. D. Kent, H. Ohno, Current-induced torques in magnetic materials. *Nat. Mater.* **11,** 372-381 (2012).
3. B. G. Park, J. Wunderlich, X. Martí, V. Holý, Y. Kurosaki, M. Yamada, H. Yamamoto, A. Nishide, J. Hayakawa, H. Takahashi, A. B. Shick, T. Jungwirth, A spin-valve-like magnetoresistance of an antiferromagnet-based tunnel junction. *Nat. Mater.* **10,** 347-351 (2011).

4. M. N. Baibich, J. M. Broto, A. Fert, F. Nguyen Van Dau, F. Petroff, P. Etienne, G. Creuzet, A. Friederich, and J. Chazelas, Giant magnetoresistance of (001)Fe/(001)Cr magnetic superlattices. *Phys. Rev. Lett.* **61,** 2472-2475 (1988).

5. Q. L. He, X. Kou, A. J. Grutter, G. Yin, L. Pan, X. Che, Y. Liu, T. Nie, B. Zhang, S. M. Disseler, B. J. Kirby, W. Ratcliff II, Q. Shao, K. Murata, X. Zhu, G. Yu, Y. Fan, M. Montazeri, X. Han, J. A. Borchers, K. L. Wang, Tailoring exchange couplings in magnetic topological-insulator/ antiferromagnet heterostructures. *Nat. Mater.* **16,** 94-100 (2017).

6. C.-Z. Chang, J. Zhang, X. Feng, J. Shen, Z. Zhang, M. Guo, K. Li, Y. Ou, P. Wei, L.-L. Wang, Z.-Q. Ji, Y. Feng, S. Ji, X. Chen, J. Jia, X. Dai, Z. Fang, S.-C. Zhang, K. He, Y. Wang, L. Lu, X.-C. Ma, Q.-K. Xue, Experimental observation of the quantum anomalous Hall effect in a magnetic topological insulator. *Science* **340,** 167-170 (2013).

7. L. A. Wray, S.-Y. Xu, Y. Xia, D. Hsieh, A. V. Fedorov, Y. S. Hor, R. J. Cava, A. Bansil, H. Lin, M. Z. Hasan, A topological insulator surface under strong Coulomb, magnetic and disorder perturbations. *Nat. Phys.* **7,** 32-37 (2011).

8. M. Mogi, M. Kawamura, R. Yoshimi, A. Tsukazaki, Y. Kozuka, N. Shirakawa, K. S. Takahashi, M. Kawasaki, Y. Tokura, A magnetic heterostructure of topological insulators as a candidate for an axion insulator. *Nat. Mater.* **16,** 516-521 (2017).

9. L. Šmejkal, Y. Mokrousov, B. Yan, A. H. MacDonald, Topological antiferromagnetic spintronics. *Nat. Phys.* **14,** 242-251 (2018).

10. M. M. Otrokov, I. I. Klimovskikh, H. Bentmann, A. Zeugner, Z. S. Aliev, S. Gass, A. U. B. Wolter, A. V. Koroleva, D. Estyunin, A. M. Shikin, M. Blanco-Rey, M. Hoffmann, A. Y. Vyazovskaya, S. V. Eremeev, Y. M. Koroteev, I. R. Amiraslanov, M. B. Babanly, N. T. Mamedov, N. A. Abdullayev, V. N. Zverev, B. Buchner, E. F. Schwier, S. Kumar, A. Kimura, L. Petaccia, G. D. Santo, R. C. Vidal, S. Schatz, K. Kißner, C.-H. Min, S. K. Moser, T. R. F. Peixoto, F. Reinert, A. Ernst, P. M. Echenique, A. Isaeva, E. V. Chulkov, Prediction and observation of the first antiferromagnetic topological insulator. arXiv:1809.07389 [cond-mat.mtrl-sci] (19 September 2018).

11. Y. Gong, J. Guo, J. Li, K. Zhu, M. Liao, X. Liu, Q. Zhang, L. Gu, L. Tang, X. Feng, D. Zhang, W. Li, C. L. Song, L. L. Wang, P. Yu, X. Chen, Y. Y. Wang, H. Yao, W. H. Wen, Y. Xu, S.-C. Zhang, X. C. Ma, Q.-K. Xue, K. He, Experimental realization of an intrinsic magnetic topological insulator. *Chinese Phys. Lett.* **36**, 076801 (2019).

12. D. Zhang, M. Shi, K. He, D. Xing, H. Zhang, J. Wang, Topological axion states in magnetic insulator $MnBi_2Te_4$ with the quantized magnetoelectric effect. *Phys. Rev. Lett.* **122**, 206401 (2019).

13. J. Li, Y. Li, S. Du, Z. Wang, B.-L. Gu, S.-C. Zhang, K. He, W. Duan, Y. Xu, Intrinsic magnetic topological insulators in van der Waals layered $MnBi_2Te_4$-family materials. *Sci. Adv.* **5**, eaaw5685 (2019).

14. X-L. Qi, T. L. Hughes, S-C. Zhang, Topological field theory of time-reversal invariant insulators. *Phys. Rev. B* **78,** 195424 (2008).

15. M. M. Otrokov, T. V. Menshchikova, M. G. Vergniory, I. P. Rusinov, A. Y. Vyazovskaya, Y. M. Koroteev, G. Bihlmayer, A. Ernst, P. M. Echenique, A. Arnau, E. V. Chulkov, Highly-ordered wide bandgap materials for quantized anomalous Hall and magnetoelectric effects. *2D Mater.* **4**, 025082 (2017).

16. K. S. Novoselov, A. Mishchenko, A. Carvalho, A. H. C. Neto, 2D materials and van der Waals heterostructures. *Science* **353,** aac9439 (2016).

17. Z. S. Aliev, I. R. Amiraslanov, D. I. Nasonova, A. V. Shevelkov, N. A. Abdullayev, Z. A. Jahangirli, E. N. Orujlu, M. M. Otrokov, N. T. Mamedov, M. B. Babanly, E. V. Chulkov, Novel ternary layered manganese bismuth tellurides of the $MnTe-Bi_2Te_3$ system: Synthesis and crystal structure. *J. Alloys Compd.* **789**, 443-450 (2019).

18. A. Zeugner, F. Nietschke, A. U. B. Wolter, S. Gaß, R. C. Vidal, T. R. F. Peixoto, D. Pohl, C. Damm, A. Lubk, R. Hentrich, S. K. Moser, C. Fornari, C. H. Min, S. Schatz, K. Kißner, M. Ünzelmann, M. Kaiser, F. Scaravaggi, B. Rellinghaus, K. Nielsch, C. Hess, B. Büchner, F. Reinert, H. Bentmann, O. Oeckler, T. Doert, M. Ruck, A. Isaeva, Chemical aspects of the antiferromagnetic topological insulator $MnBi_2Te_4$. *Chem. Mater.* **31**, 2795-2806 (2019).

19. S. H. Lee, Y. Zhu, Y. Wang, L. Miao, T. Pillsbury, S. Kempinger, D. Graf, N. Alem, C.-Z. Chang, N. Samarth, Z. Mao, Spin scattering and noncollinear spin structure-induced intrinsic anomalous Hall effect in antiferromagnetic topological insulator $MnBi_2Te_4$. arXiv:1812.00339 [cond-mat.mtrl-sci] (2 December 2018).

20. J.-Q. Yan, Q. Zhang, T. Heitmann, Z. Huang, K. Y. Chen, J.-G. Cheng, W. Wu, D. Vaknin, B. C. Sales, R. J. McQueeney, Crystal growth and magnetic structure of $MnBi_2Te_4$. *Phys. Rev. Materials* **3**, 064202 (2019).

21. J. Cui, M. Shi, H. Wang, F. Yu, T. Wu, X. Luo, J. Ying, X. Chen, Transport properties of thin flakes of the antiferromagnetic topological insulator $MnBi_2Te_4$. *Phys. Rev. B* **99**, 155125 (2019).

22. M. Z. Hasan, C. L. Kane, Colloquium: Topological insulators. *Rev. Mod. Phys.* **82**, 3045 (2010).

23. R. S. K. Mong, A. M. Essin, J. E. Moore, Antiferromagnetic topological insulators. *Phys. Rev. B* **81**, 245209 (2010).

24. A. A. Mostofi, J. R. Yates, G. Pizzi, Y. S. Lee, I. Souza, D. Vanderbilt, N. Marzari, An updated version of wannier90: A tool for obtaining maximally-localised Wannier functions.

*Comput. Phys. Commun.* **185**, 2309-2310 (2014).

25. Q. Wu, S. Zhang, H.-F. Song, M. Troyer, A. A. Soluyanov, WannierTools: An open-source software package for novel topological materials. *Comput. Phys. Commun.* **224**, 405 (2018).

26. Y. L. Chen, J. G. Analytis, J.-H. Chu, Z. K. Liu, S.-K. Mo, X. L. Qi, H. J. Zhang, D. H. Lu, X. Dai, Z. Fang, S. C. Zhang, I. R. Fisher, Z. Hussain, Z.-X. Shen, Experimental realization of a three-dimensional topological insulator, $Bi_2Te_3$. *Science* **325,** 178-181 (2009).

27. S. V. Eremeev, G. Landolt, T. V. Menshchikova, B. Slomski, Y. M. Koroteev, Z. S. Aliev, M. B. Babanly, J. Henk, A. Ernst, L. Patthey, A. Eich, A. A. Khajetoorians, J. Hagemeister, O. Pietzsch, J. Wiebe, R. Wiesendanger, P. M. Echenique, S. S. Tsirkin, I. R. Amiraslanov, J. H. Dil, E. V. Chulkov, Atom-specific spin mapping and buried topological states in a homologous series of topological insulators. *Nat. Commun.* **3,** 635 (2012).

28. J. Sánchez-Barriga, A. Varykhalov, G. Springholz, H. Steiner, R. Kirchschlager, G. Bauer, O. Caha, E. Schierle, E. Weschke, A. A. Ünal, S. Valencia, M. Dunst, J. Braun, H. Ebert, J. Minár, E. Golias, L. V. Yashina, A. Ney, V. Holý, O. Rader, Nonmagnetic band gap at the Dirac point of the magnetic topological insulator $(Bi_{1-x}Mn_x)_2Se_3$. *Nat. Commun.* **7**, 10559 (2016).

29. C.-Z. Chang, P. Tang, Y.-L. Wang, X. Feng, K. Li, Z. Zhang, Y. Wang, L.-L. Wang, X. Chen, C. Liu, W. Duan, K. He, X.-C. Ma, Q.-K. Xue, Chemical-potential-dependent gap opening at the Dirac surface states of $Bi_2Se_3$ induced by aggregated substitutional Cr atoms. *Phys. Rev. Lett.* **112**, 056801 (2014).

30. Y. L. Chen, J.-H. Chu, J. G. Analytis, Z. K. Liu, K. Igarashi, H.-H. Kuo, X. L. Qi, S. K. Mo, R. G. Moore, D. H. Lu, M. Hashimoto, T. Sasagawa, S. C. Zhang, I. R. Fisher, Z. Hussain, Z. X. Shen, Massive dirac fermion on the surface of a magnetically doped topological insulator. *Science* **329,** 659-662 (2010).

31. S.-Y. Xu, M. Neupane, C. Liu, D. Zhang, A. Richardella, L. A. Wray, N. Alidoust, M. Leandersson, T. Balasubramanian, J. Sánchez-Barriga, O. Rader, G. Landolt, B. Slomski, J. H. Dil, J. Osterwalder, T.-R. Chang, H.-T. Jeng, H. Lin, A. Bansil, N. Samarth, M. Z. Hasan, Hedgehog spin texture and Berry's phase tuning in a magnetic topological insulator. *Nat. Phys.* **8,** 616-622 (2012).

32. T. Okuda, T. Maegawa, M. Ye, K. Shirai, T. Warashina, K. Miyamoto, K. Kuroda, M. Arita, Z. S. Aliev, I. R. Amiraslanov, M. B. Babanly, E. V. Chulkov, S. V. Eremeev, A. Kimura, H. Namatame, M. Taniguchi, Experimental evidence of hidden topological surface states in $PbBi_4Te_7$. *Phys. Rev. Lett.* **111,** 206803 (2013).

33. N. Nagaosa, J. Sinova, S. Onoda, A. H. MacDonald, N. P. Ong, Anomalous hall effect. *Rev. Mod. Phys.* **82,** 1539-1592 (2010).

34. S. Onoda, N. Sugimoto, N. Nagaosa, Intrinsic versus extrinsic anomalous Hall effect in ferromagnets. *Phys. Rev. Lett.* **97,** 126602 (2006).

35. M. Lee, Y. Onose, Y. Tokura, N. P. Ong, Hidden constant in the anomalous Hall effect of high-purity magnet MnSi. *Phys. Rev. B* **75,** 172403 (2007).

36. E. Liu, Y. Sun, N. Kumar, L. Muechler, A. Sun, L. Jiao, S.-Y. Yang, D. Liu, A. Liang, Q. Xu, J. Kroder, V. Süß, H. Borrmann, C. Shekhar, Z. Wang, C. Xi, W. Wang, W. Schnelle, S. Wirth, Y. Chen, S. T. B. Goennenwein, C. Felser, Giant anomalous Hall effect in a ferromagnetic kagome-lattice semimetal. *Nat. Phys.* **14,** 1125-1131 (2018).

37. H. Chen, Q. Niu, A. H. MacDonald, Anomalous Hall effect arising from noncollinear antiferromagnetism. *Phys. Rev. Lett.* **112,** 017205 (2014).

38. T. Jungwirth, W. A. Atkinson, B. H. Lee, A. H. MacDonald, Interlayer coupling in ferromagnetic semiconductor superlattices. *Phys. Rev. B* **59,** 9818-9821 (1999).

39. T. Moriya, Anisotropic superexchange interaction and weak ferromagnetism. *Phys. Rev.* **120,** 91-98 (1960).

40. R. Yu, W. Zhang, H.-J. Zhang, S.-C. Zhang, X. Dai, Z. Fang, Quantized anomalous Hall effect in magnetic topological insulators. *Science* **329,** 61-64 (2010).

41. M. M. Otrokov, I. P. Rusinov, M. Blanco-Rey, M. Hoffmann, A. Y. Vyazovskaya, S. V. Eremeev, A. Ernst, P. M. Echenique, A. Arnau, E. V. Chulkov, Unique thickness-dependent properties of the van der Waals interlayer antiferromagnet $MnBi_2Te_4$ films. *Phys. Rev. Lett.* **122**, 107202 (2018).

42. B. H. Toby, EXPGUI, a graphical user interface for GSAS. *J. Appl. Cryst.* **34**, 210-213 (2001).

43. G. Kresse, J. Furthmuller, Efficient iterative schemes for ab initio total-energy calculations using a plane-wave basis set. *Phys. Rev. B* **54**, 11169 (1996).

44. G. Kresse, J. Furthmuller, Efficiency of ab-initio total energy calculations for metals and semiconductors using a plane-wave basis set. *Comput. Mater. Sci.* **6**, 15-50 (1996).

45. J. Heyd, G. E. Scuseria, M. Ernzerhof, Assessment and validation of a screened Coulomb hybrid density functional. *J. Chem. Phys.* **120**, 7274 (2004).

46. A. V. Krukau, O. A. Vydrov, A. F. Izmaylov, G. E. Scuseria, Influence of the exchange screening parameter on the performance of screened hybrid functionals. *J. Chem. Phys.* **125**, 224106 (2006).

**Acknowledgements:** The authors would like to thank Y. Sato (Tokyo Institute of Technology) for her technical support in the Auger electron spectroscopy measurements. J. W. would like to thank Prof. S. Matsuishi (Tokyo Institute of Technology) for valuable discussions. **Funding:** This work was supported by the Japan Society for the Promotion of Science (JSPS) through a Grant-in-Aid for Scientific Research (S), No.17H06153 and by the MEXT Element Strategy

Initiative to form a research core. **Author contributions:** H. H. proposed the idea behind the research and supervised the project. J. W. performed the materials synthesis and characterization. M. S. performed the STEM measurements. J. W. and F. L. performed the physical properties measurements (above 2 K) and data analysis. K. I. and S. O. performed the electrical transport properties measurements below 2 K and data analysis. Y. O., R. Y., K. H., and H. K. performed the ARPES measurements and data analysis. T. I. performed theoretical calculations. J. W., F. L. and H. H. co-wrote the paper. All authors discussed the results and commented on and approved the manuscript. **Competing interests:** The authors declare no competing financial interests. **Data and materials availability:** All data needed to evaluate the conclusions in the paper are present in the paper and the supplementary materials. Additional data related to this paper may be requested from the authors.

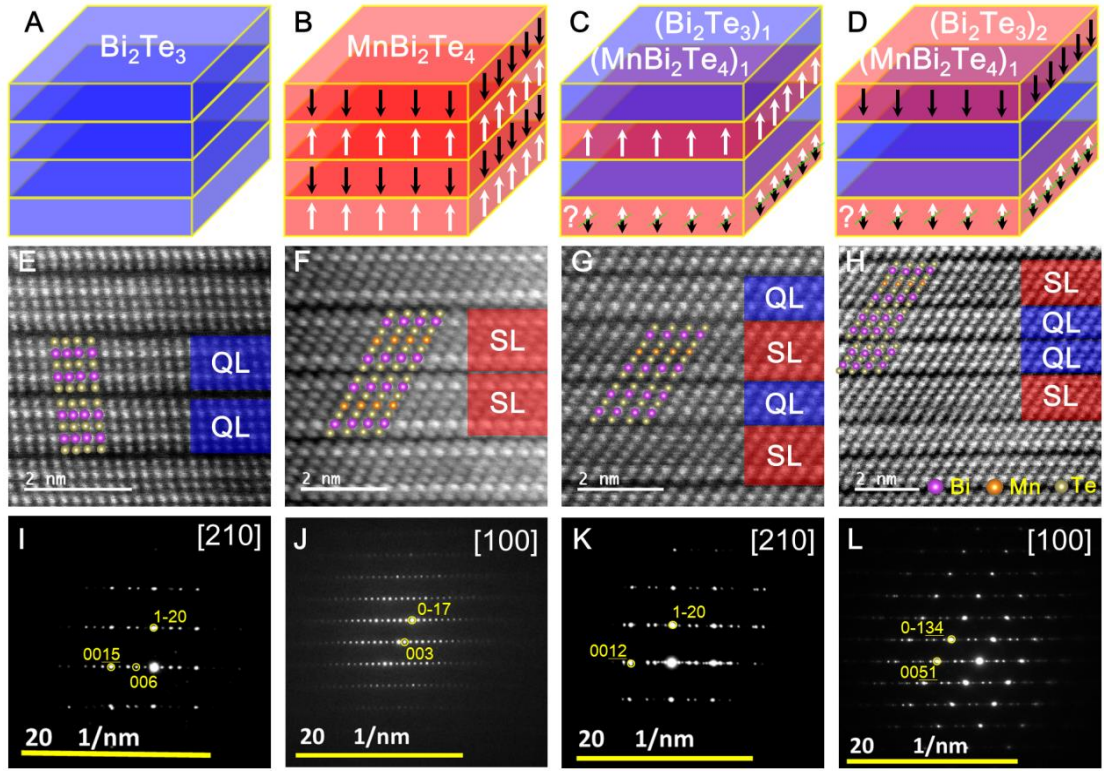

**Fig. 1. Magnetic van der Waals heterostructures of (MnBi$_2$Te$_4$)$_m$(Bi$_2$Te$_3$)$_n$.** (**A-D**) Schematics of the evolution of the heterostructures. The arrows show the spin orientation of Mn with black pointing down and white pointing up. The question marks in **C** and **D** show the uncertainty of the spin orientations due to complex magnetic interactions. (**E-H**) Atomic resolution HAADF-STEM images of the compounds displayed in **A-D**. The images are taken along a zone axis perpendicular to the *c* axis. QL stands for quintuple layer and SL stands for septuple layer. (**I-L**) Selected area electron diffraction (SAED) patterns of the compounds displayed in **A-D**.

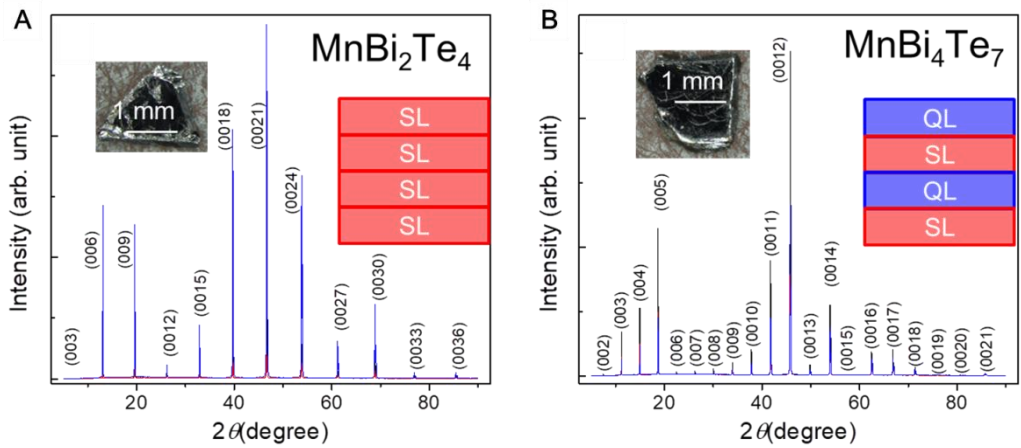

**Fig. 2. XRD patterns of single crystals.** **(A)** MnBi$_2$Te$_4$. **(B)** MnBi$_4$Te$_7$. The measurement was performed on single-crystalline pieces (shown in the insets) with only *ab* plane exposed to X-ray. The insets also show the structure models based on SL and QL van der Waals layers.

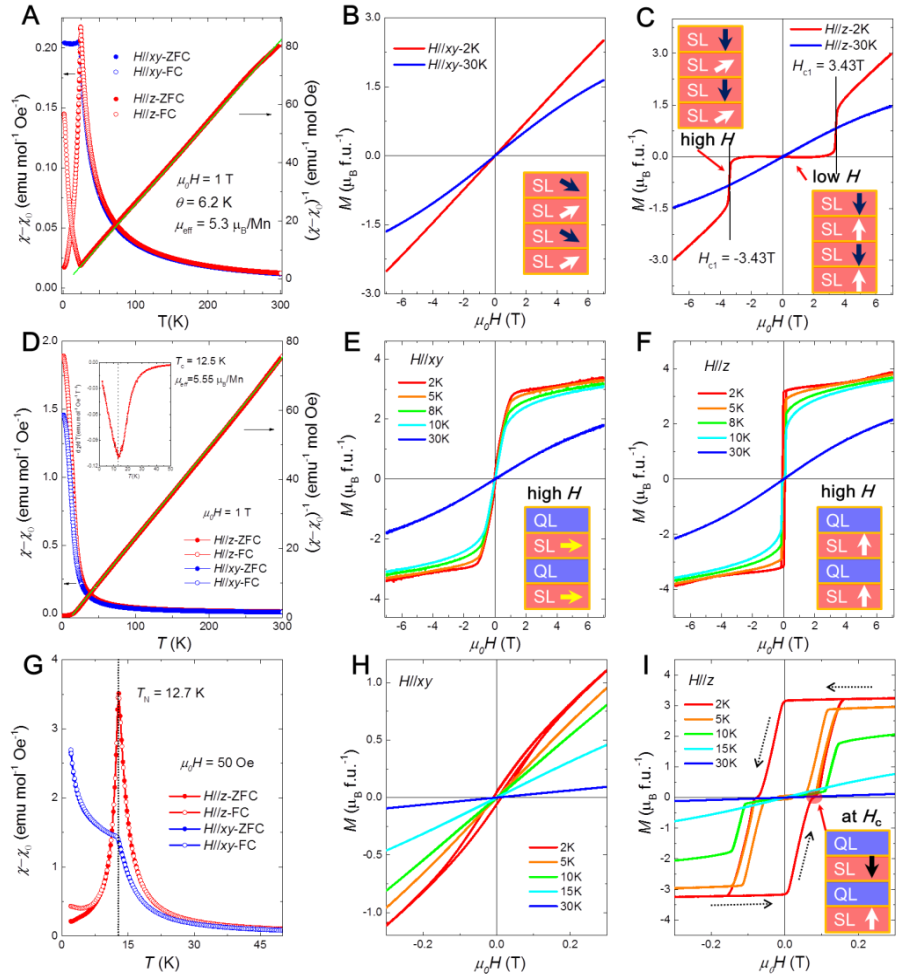

**Fig. 3. Magnetic properties of MnBi$_2$Te$_4$ and MnBi$_4$Te$_7$ single crystals.** (**A-C**) Magnetic susceptibility and magnetization of MnBi$_2$Te$_4$. The parameters θ and $\mu_{\text{eff}}$ are the Curie-Weiss temperature and effective moment, respectively. (**D-F**) Magnetic susceptibility and magnetization of MnBi$_4$Te$_7$ at high fields. (**G-I**) Magnetic susceptibility and magnetization of MnBi$_4$Te$_7$ at low fields. The black arrows with dotted lines in **I** show the sweep directions of the magnetic field. The heterostructures and spin structures are schematically shown as insets in **B**, **C, E, F** and **I**.

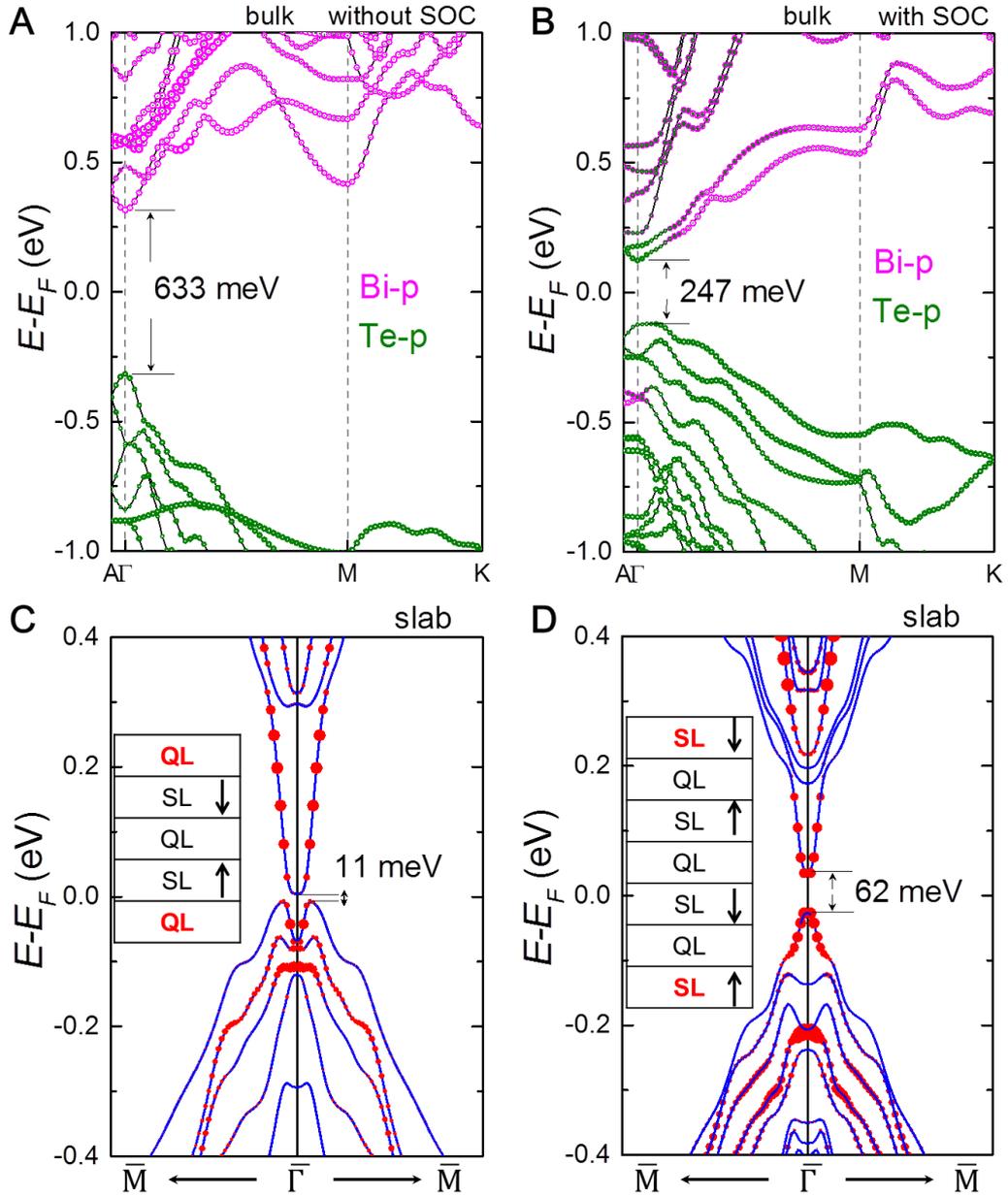

**Fig. 4. DFT band structures of MnBi$_4$Te$_7$.** (**A**) Bulk band structure without SOC. (**B**) Bulk band structure with SOC. (**C**) Band structure of a QL-terminated five-van der Waals-layer slab. (**D**) Band structure of a SL-terminated seven-van der Waals-layer slab. The calculations were performed assuming an AFM ground state. The thickness of the band is proportional to the contribution of the indicated atoms (panel A and B) or van der Waals layers (QL/SL in panel C and D).

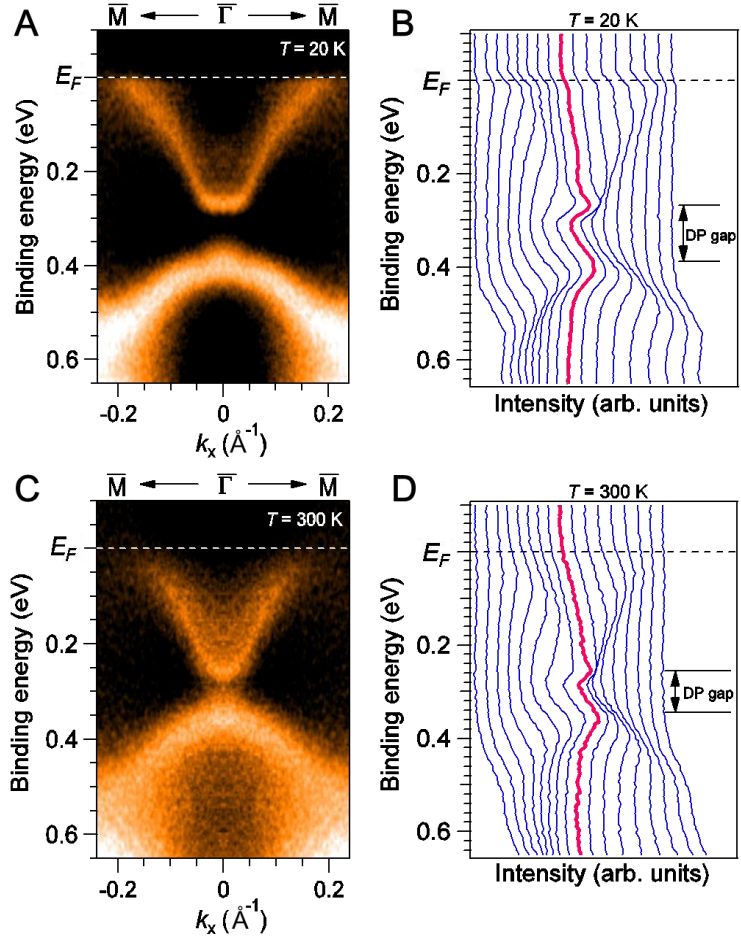

**Fig. 5. Surface band structure of MnBi$_4$Te$_7$ at a photon energy of 48 eV.** (**A**, **C**) Measured SS along $\bar{\Gamma} - \bar{M}$ direction at 20 K and 300 K, respectively. The intensity plots are symmetrized with respect to the center lines and averaged. (**B, D**) The energy distribution curves (EDCs) extracted from the intensity maps of **A** and **C**, respectively, in the range of –0.24 Å$^{-1}$ < $k_x$ < 0.24 Å$^{-1}$. The intensity at the $\bar{\Gamma}$ point is shown in red.

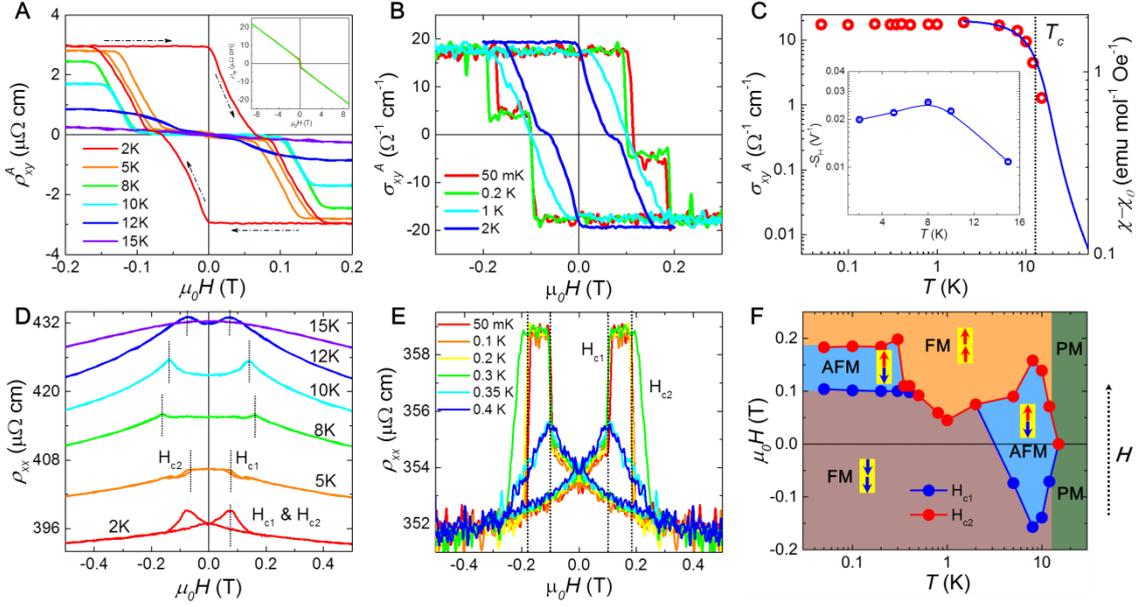

**Fig. 6. Anomalous electrical transport properties and magnetic structures of MnBi$_4$Te$_7$ single crystals.** (**A**) AH resistivity. The black arrows with dotted lines show the sweep directions of the magnetic field. The inset shows the total Hall resistivity at 2 K. (**B**) AH conductivity below 2 K. The central-symmetric data including background noise are due to the analysis method (Supplementary method). (**C**) Temperature-dependence of AH conductivity and magnetic susceptibility measured at 1 T. The inset shows the temperature-dependence of the scaling factor. (**D**) Field- and temperature-dependent magnetoresistance above 2 K. (**E**) Field and temperature dependent magnetoresistance below 1 K. (**F**) Temperature- and field-dependence of magnetic structures when field is swept from a negative field to a positive field. It is noted that measurements above 2 K and below 1 K are on two pieces of samples. The magnetic field is applied along the *c* axis for all the transport properties measurements.

# Supplementary Materials for

# Natural van der Waals Heterostructural Single Crystals with both Magnetic and Topological Properties

Jiazhen Wu[1], Fucai Liu[2,1*], Masato Sasase[1], Koichiro Ienaga[3], Yukiko Obata[4], Ryu Yukawa[4], Koji Horiba[4], Hiroshi Kumigashira[5,4], Satoshi Okuma[3], Takeshi Inoshita[1,6], and Hideo Hosono[1,7*]

[1] Materials Research Center for Element Strategy, Tokyo Institute of Technology, 4259 Nagatsuta, Midori-ku, Yokohama 226-8503, Japan

[2] School of Optoelectronic Science and Engineering, University of Electronic Science and Technology of China, Chengdu, 610054 China

[3] Department of Physics, Tokyo Institute of Technology, 2-12-1 Ohokayama, Meguro-ku, Tokyo 152-8551 Japan

[4] Photon Factory and Condensed Matter Research Center, Institute of Materials Structure Science, High Energy Accelerator Research Organization (KEK), Tsukuba 305-0801, Japan

[5] Institute of Multidisciplinary Research for Advanced Materials (IMRAM), Tohoku University, Sendai 980-8577, Japan

[6] National Institute for Materials Science, Tsukuba, Ibaraki 305-0044, Japan

[7] Laboratory for Materials and Structures, Institute of Innovative Research, Tokyo Institute of Technology, 4259 Nagatsuta, Midori-ku, Yokohama 226-8503, Japan

*e-mail: fucailiu@uestc.edu.cn, hosono@msl.titech.ac.jp

**Supplementary Text:**

**Section S1. Data analysis methods for electrical transport measurements**

**Separation of Hall resistivity and longitudinal resistivity**: As the electrodes prepared for electrical transport properties measurements are not perfectly aligned for both Hall resistance ($\rho_{xy}$) and longitudinal resistance ($\rho_{xx}$) measurements, the measured field-dependent Hall resistance ($\rho_{xy}^{\text{measure}}$) contains the contributions from not only the Hall effect but also magnetoresistance effect, and similarly the measured field-dependent longitudinal resistance ($\rho_{xx}^{\text{measure}}$) contains the contributions from both the Hall effect and magnetoresistance effect. Therefore we have, $\rho_{xy}^{\text{measure}} = \rho_{xy}^{\text{H}} + \rho_{xx}^{\text{H}}$ and $\rho_{xx}^{\text{measure}} = \rho_{xx}^{\text{L}} + \rho_{xy}^{\text{L}}$, where $\rho_{xy}^{\text{H}}$ and $\rho_{xy}^{\text{L}}$ are contributed from the Hall effect in Hall resistance and longitudinal resistance measurements, respectively, and $\rho_{xx}^{\text{H}}$ and $\rho_{xx}^{\text{L}}$ are contributed from the magnetoresistance effect in Hall resistance and longitudinal resistance measurements, respectively. In order to separate the two contributions, we analyzed the data based on two equations: $\rho_{xy}(H) = -\rho_{xy}(-H)$ and $\rho_{xx}(H) = \rho_{xx}(-H)$. The pure Hall resistivity was obtained by $\rho_{xy}^{\text{H}}(H) = (\rho_{xy}^{\text{measure}}(H) - \rho_{xy}^{\text{measure}}(-H))/2$, and the pure longitudinal resistivity was obtained by $\rho_{xx}^{\text{L}}(H) = (\rho_{xx}^{\text{measure}}(H) + \rho_{xy}^{\text{measure}}(-H))/2$. To analyze the data in such a way, we measured both $\rho_{xy}^{\text{measure}}$ and $\rho_{xx}^{\text{measure}}$ with magnetic field swept from a negative field to positive field and then from a positive field to a negative field. It should be noted that, by using such an analytical method, the data were symmetrized/ antisymmetrized even including the background noise. The ordinary Hall resistivity ($\rho_{xy}^{N}$) and AH resistivity ($\rho_{xy}^{A}$) were separated based on $\rho_{xy} = \rho_{xy}^{N} + \rho_{xy}^{A} = R_H B + \rho_{xy}^{A}(B)$, where $R_H$ is the ordinary Hall coefficient and $\rho_{xy}^{N} = R_H B$. At high fields (larger than 0.2 T for MnBi$_4$Te$_7$), the expression can be simplified as $\rho_{xy} = R_H B + C$, where C is a constant value of the saturated AH resistivity. Therefore, a linear fitting of the data of $\rho_{xy}$ at high fields will give rise to $R_H$ and C. Similarly, we can separate the ordinary Hall conductivity and AH conductivity from the total Hall conductivity.

**Separation of ordinary Hall conductivity and anomalous Hall conductivity**: Hall conductivity can be converted according to $\sigma_{xy} = \rho_{yx}/(\rho_{xx}^2 + \rho_{yx}^2) \cong \rho_{yx}/\rho_{xx}^2 = \rho_{yx}^{N}/\rho_{xx}^2 + \rho_{yx}^{A}/\rho_{xx}^2$, where $\sigma_{xy}^{N} = \rho_{yx}^{N}/\rho_{xx}^2$ is the ordinary Hall conductivity due to the Lorentz force that the external field applied to the conduction carriers and $\sigma_{xy}^{A} = \rho_{yx}^{A}/\rho_{xx}^2$ is the anomalous Hall conductivity. $\sigma_{xy}^{N}$ can be obtained by linear fittings of the field-dependent $\sigma_{xy}$ at high fields where $\sigma_{xy}^{A}$ is saturated. Here, $\sigma_{xy}^{N} = \rho_{yx}^{N}/\rho_{xx}^2 = R_H \mu_0 H/\rho_{xx}^2$, in which $R_H$ is the ordinary Hall coefficient.

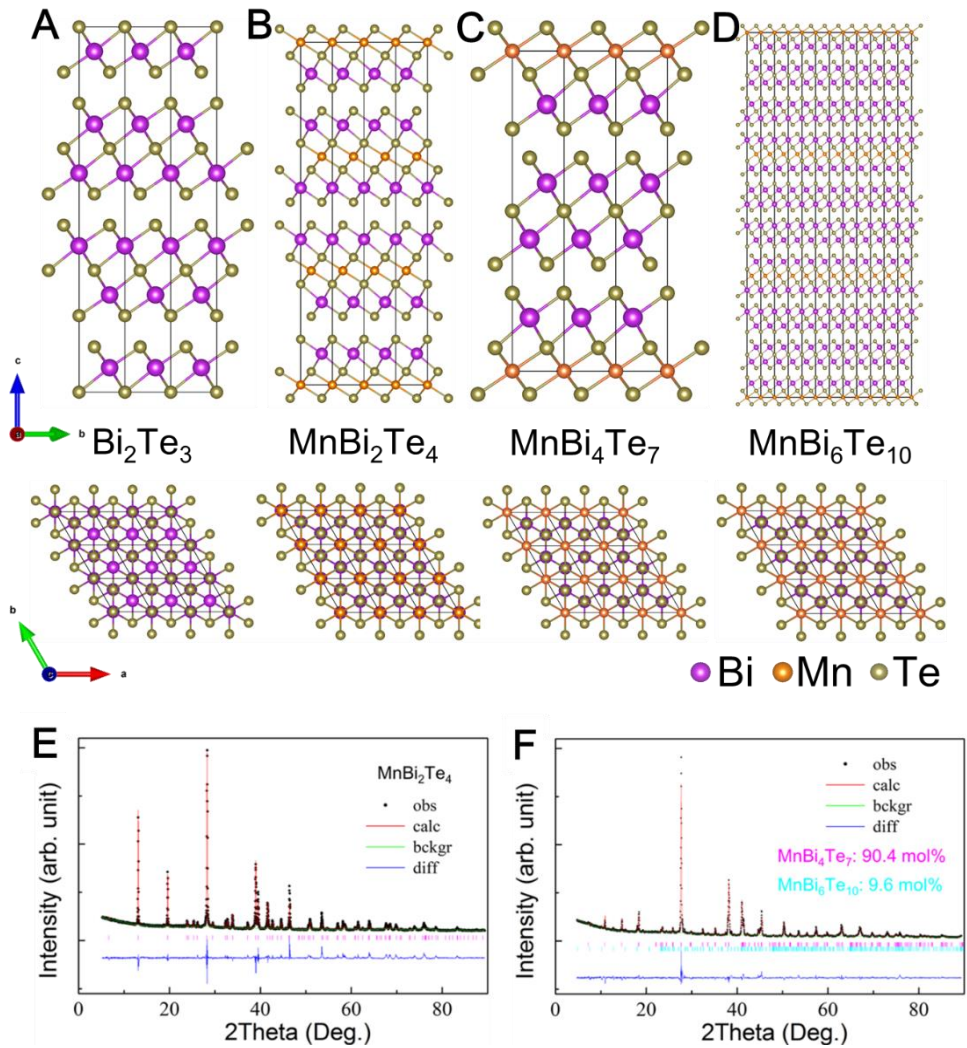

**Fig. S1. Crystal structures and XRD patterns.** (**A**)-(**D**) Crystal structures. (**E**)-(**F**) Powder X-ray diffraction patterns after Rietveld refinement (*42*).

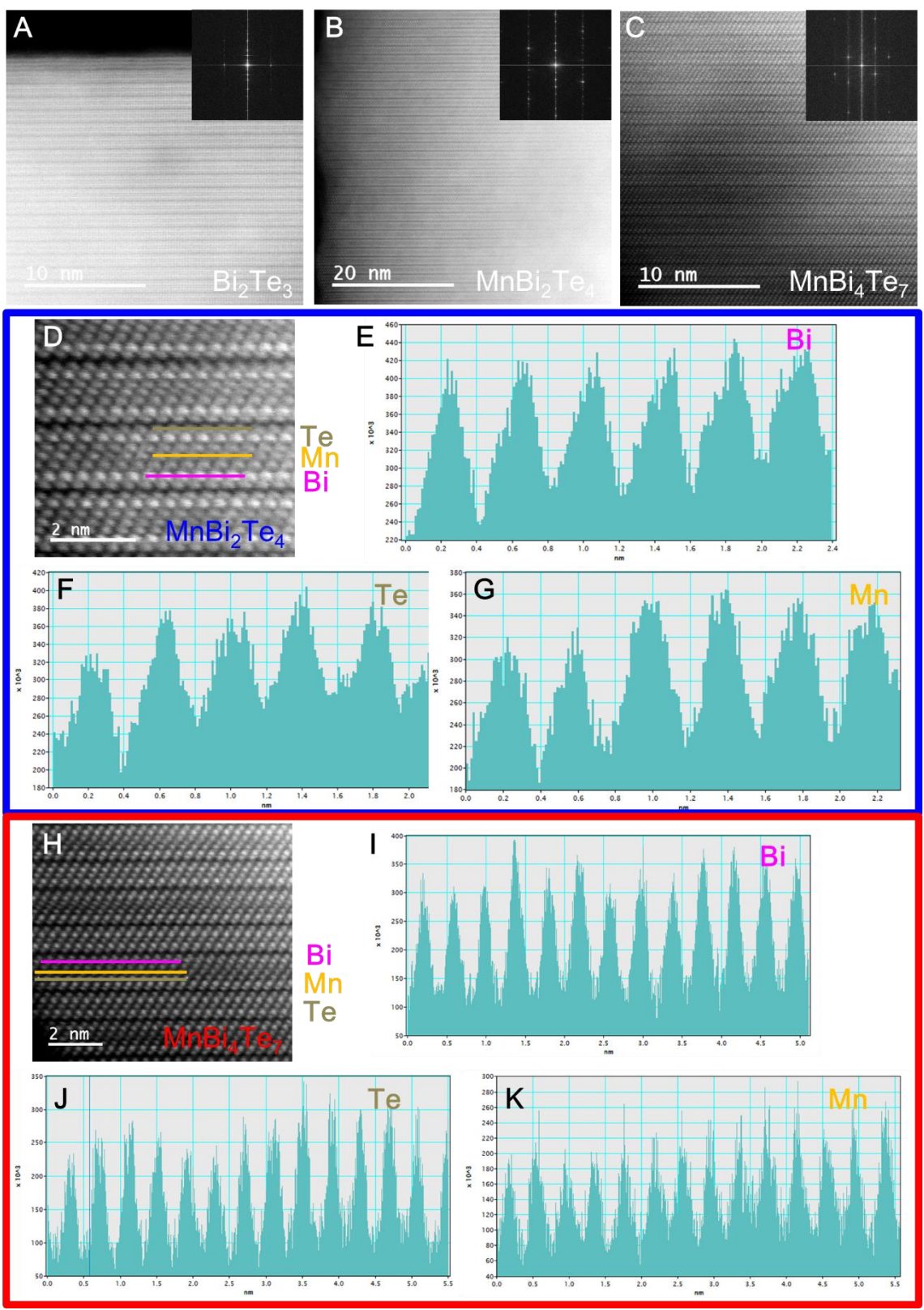

**Fig. S2. Additional STEM data.** (**A**)-(**C**) The image is taken along a zone axis perpendicular to *c* axis for $Bi_2Te_3$, $MnBi_2Te_4$ and $MnBi_4Te_7$, respectively, showing clear QL and/or SL, which are separated by van der Waals gaps. The fast Fourier transform (FFT) images are shown as insets of the figures, demonstrating the high quality of the crystal. (**D**)-(**G**) The HAADF-STEM image of MnBi$_2$Te$_4$ and the intensity profiles of Bi, Te and Mn. It is clear that the intensity follows Bi>Te>Mn, consistent with their atomic numbers. (**H**)-(**K**) The HAADF-STEM image of MnBi$_4$Te$_7$ and the intensity profiles of Bi, Te and Mn. The intensity also follows Bi>Te>Mn.

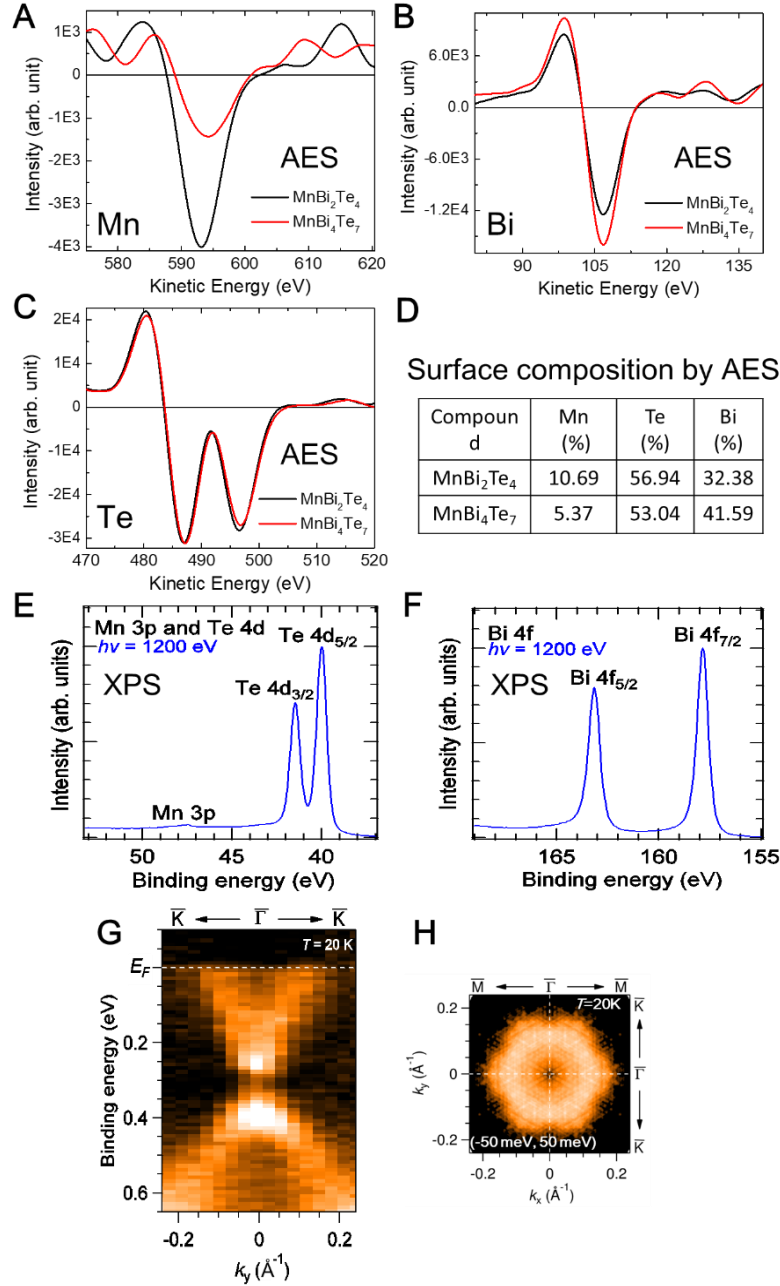

**Fig. S3. Additional surface characterization.** (**A**)-(**C**) Auger electron spectra (AES) of MnBi$_2$Te$_4$ and MnBi$_4$Te$_7$ single crystals for Mn (**A**), Bi (**B**) and Te (**C**). The spectra were taken after surface contaminations were completely removed by Ar sputtering for more than 10 minutes. (**D**) the composition ratios determined by the AES spectra. (**E**)-(**F**) X-ray photoelectron spectroscopy (XPS) measurements for MnBi$_4$Te$_7$ single crystal with a photon energy of 1200 eV. The measurements were performed on a fresh surface, which was cleaved *in situ* under an ultrahigh vacuum of $1\times10^{-10}$ Torr, in the same chamber for ARPES measurements. (**G**)-(**H**) Additional ARPES data for MnBi$_4$Te$_7$. The plots in (**H**) are symmetrized with respect to the six-fold rotational symmetric lines and averaged.

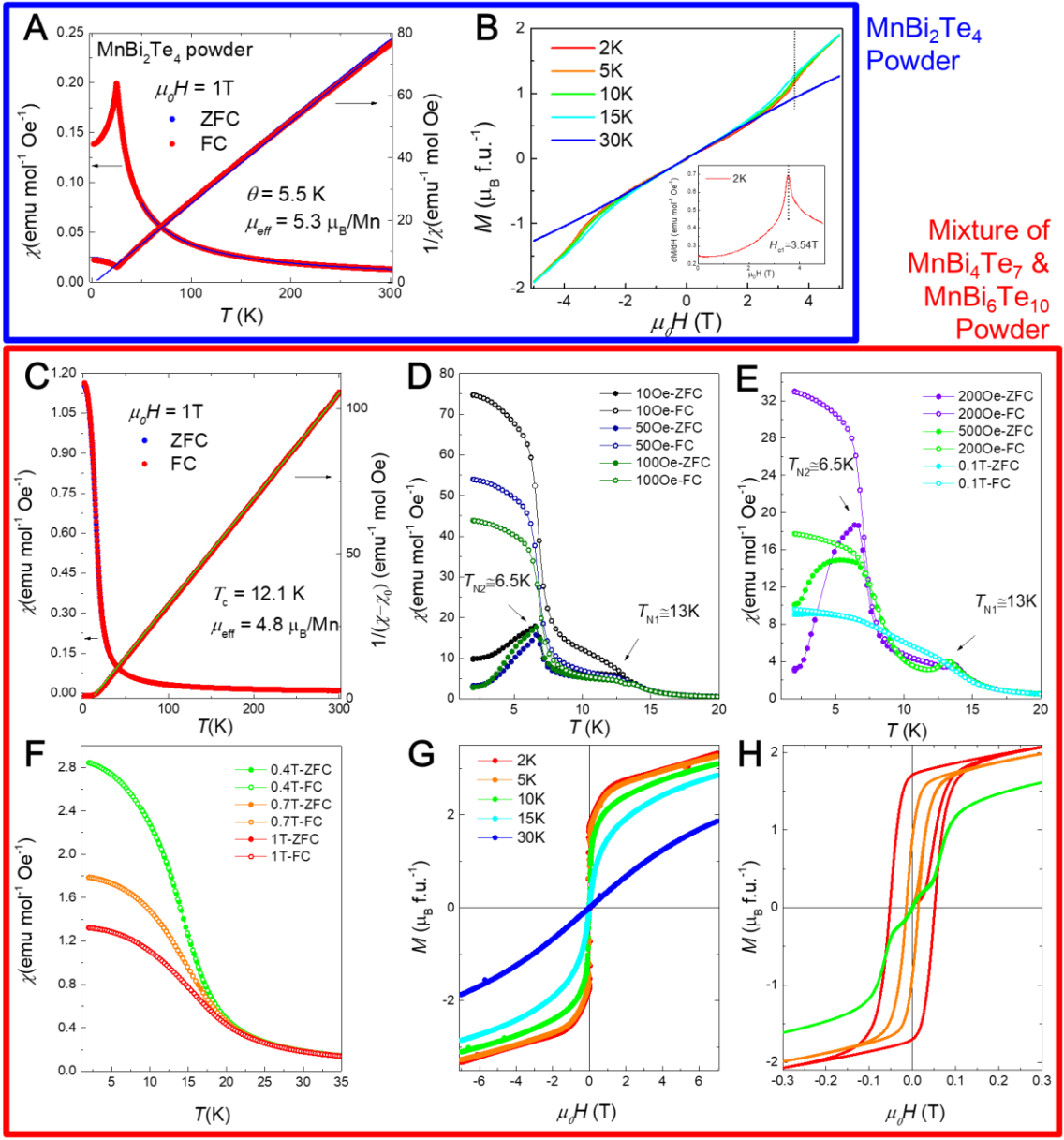

**Fig. S4. Magnetization of polycrystalline samples.** (**A**) Temperature-dependent magnetic susceptibility of $MnBi_2Te_4$. The powder sample shows an AFM transition at around 25 K, however the Curie-Weiss temperature is shown to be positive (θ = 5.5 K), indicating that the intralayer FM interactions are stronger than the interlayer AFM interactions in paramagnetic (PM) phase. This is consistent with the single crystal results shown in Fig. 3 in the main paper. The effective moment ($\mu_{eff}$) was derived to be $5.3\mu_B$/Mn, showing that Mn is divalent ($Mn^{2+}$). (**B**) Field-dependent magnetization of $MnBi_2Te_4$. There is a spin-flop transition at $|H_c| = 3.54$T at 2K. (**C**)-(**H**) Magnetization of the mixture of $MnBi_4Te_7$ and $MnBi_6Te_{10}$ (the XRD patterns are shown in Fig. S1F). (**C**) Temperature-dependent magnetic susceptibility measured at 1 T. It shows a FM transition at $T_c = 12.1\ K$. (**D**)-(**F**) Temperature-dependent magnetic susceptibility

measured at different fields from 10 Oe to 1 T. At low fields ($\mu_0 H < 0.1$ T), the magnetic transition is very complicated, showing an AFM transition ($T_{N1} \cong 13$ K) contributed from MnBi$_4$Te$_7$, and a spin glass-like transition at $T_{N2} \cong 6.5$ K. (**G**)-(**H**) Field-dependent magnetization measured at different temperatures. At low temperatures (T $\leq$ 5 K), magnetic hysteresis arise, indicating the appearance of predominant FM states even at low fields.

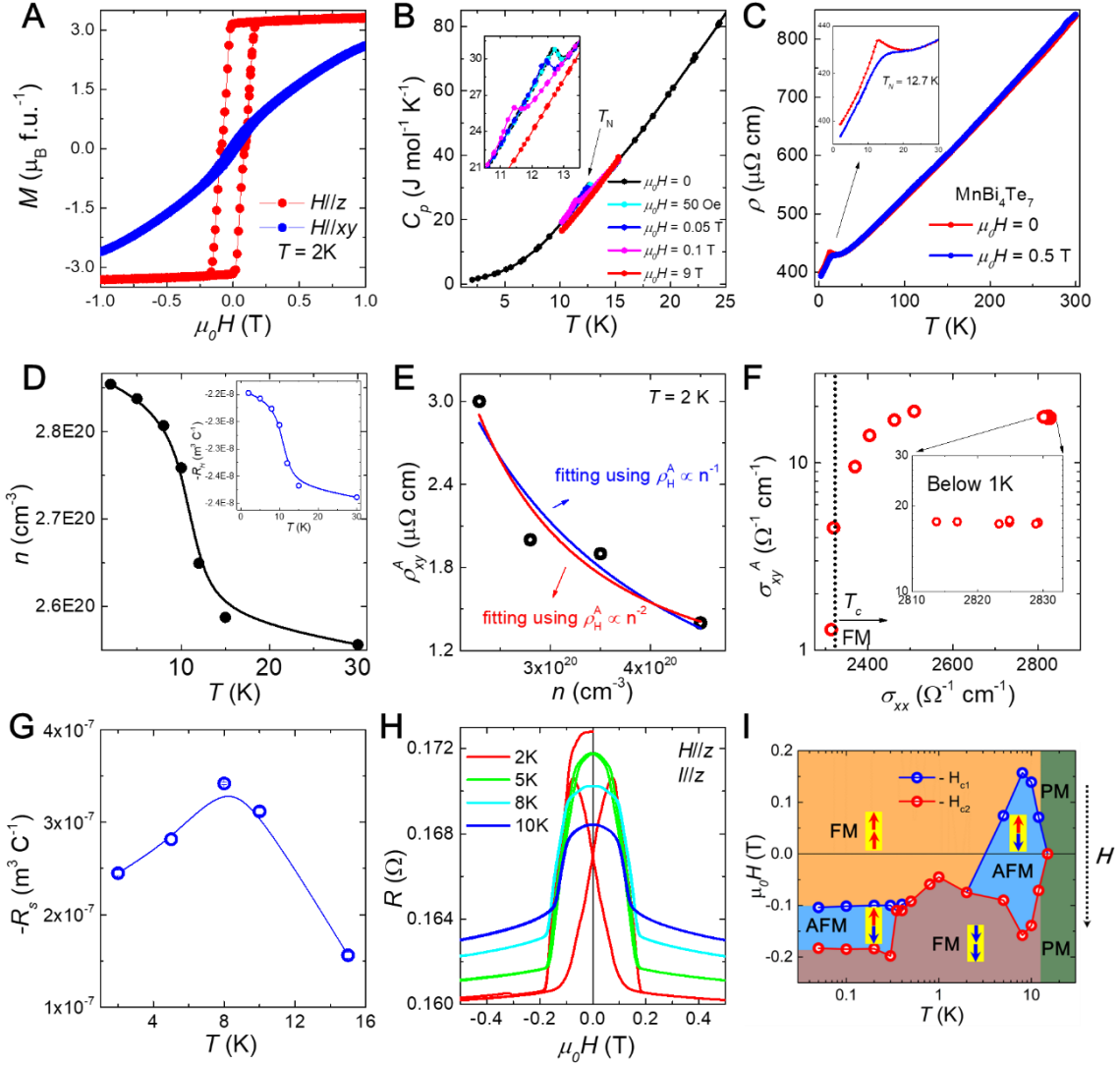

**Fig. S5. Additional physical properties of MnBi$_4$Te$_7$ single crystal.** (**A**) Field-dependent magnetization along different crystal orientations at 2 K. It indicates that *c* axis is the magnetization-easy axis. (**B**) Low temperature specific heat measured at different magnetic fields. (**C**) Resistivity. The current flowed in *a-b* plane, and the applied magnetic field was along *c* axis. (**D**) The derived carrier density from the ordinary Hall effect at high fields (Fig. 6a in the main paper). The inset shows the corresponding Hall coefficient. (**E**) Carrier density dependence of anomalous hall resistivity. The data are taken from four pieces of samples from different batches measured at 2K and 0.2 T. The red and blue lines are the fitting results using different models. It looks that $\rho_H^A \propto n^{-2}$ fits the result better than $\rho_H^A \propto n^{-1}$. This is reasonable if we write $\rho_H^A \cong \sigma_H^A \rho_{xx}^2$, where $\sigma_H^A$ is the anomalous Hall conductivity and $\rho_{xx}$ is the longitudinal resistivity, and assume that $\sigma_H^A$ is carrier density independent. Then we can derive that $\rho_H^A \propto \rho_{xx}^2 \cong \rho_0^2 = (ne\mu)^{-2}$, where $\rho_0$ is the resistivity at zero field and $\mu$ is the mobility. We can assume that $\mu$ is $n$ independent if $n$ varies only in a small range. (**F**) The relationship between the anomalous Hall conductivity and the longitudinal Hall conductivity below 15 K. $\sigma_{xy}^A$ shows a sharp drop near $T_c$, while keeps a constant value below $T_c$. The data below 1 K were taken from a different piece of sample. (**G**) Temperature dependence of the anomalous Hall coefficient ($R_s$). ($\rho_{yx}^A = R_s M$, and $R_s = S_H \rho_{xx}^2$, Here $M$ is the magnetization, $S_H = \sigma_{xy}^A/M$ is the scaling factor shown in Fig. 6c in the main paper.) (**H**) Field-dependent resistance measured at different temperatures. Both applied current and magnetic field are along $c$ axis. The measurement was performed using a two-probe method. This result is consistent with Fig. 6D in the main paper, while the magnetoresistance effect here (current flows across the magnetic and non-magnetic van der Waals layers) looks much stronger than that in Fig. 6D (current flows along van der Waals layers). (**I**) Temperature and field dependence of magnetic structures when field is swept from a positive field to a negative field. This contains the same information as in Fig. 4f in the main paper, where the field is swept from a negative field to a positive field.

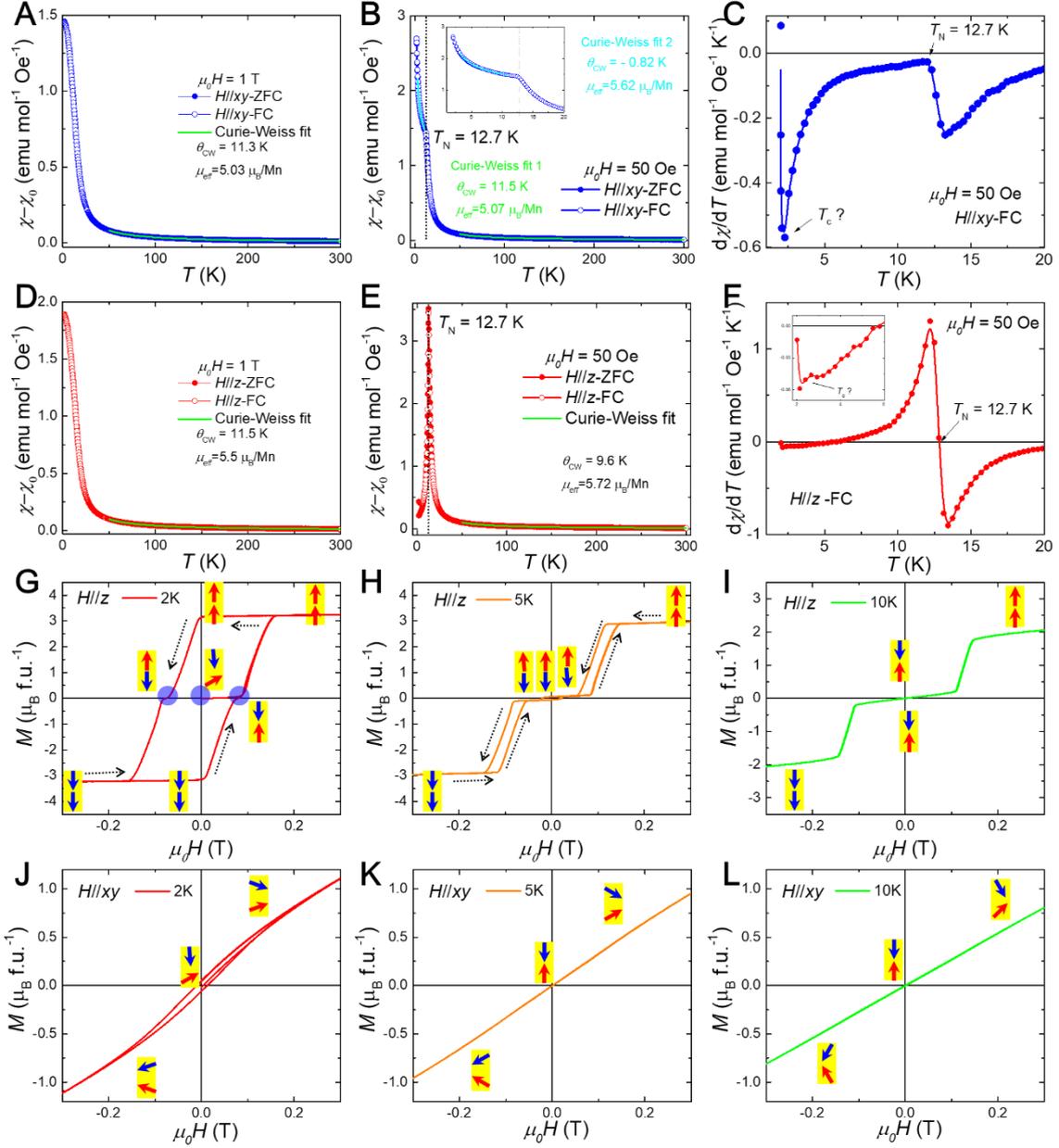

**Fig. S6. Additional information for the magnetization of MnBi$_4$Te$_7$ single crystal.** (**A**)-(**F**) Temperature- and crystal orientation-dependent magnetic measured at different fields. (**A**) and (**D**) Magnetic susceptibility measured at 1 T for $H//xy$ and $H//z$, respectively, demonstrating that the compound undergoes a FM transition at about 12 K. Curie-Weiss fitting parameters (50 K – 300 K) are shown on the figures, showing that the sample shows very weak magnetic anisotropy in the paramagnetic region. The derived effective moments ($\mu_{\text{eff}}$) show that Mn is divalent (Mn$^{2+}$). (**B**) and (**E**) Magnetic susceptibility measured at 50 Oe for $H//xy$ and $H//z$, respectively, demonstrating that the compound undergoes an AFM transition at 12.7 K ($T_N$). Below $T_N$, the interlayer AFM coupled magnetic moments are not well locked, and there is an upturn in (**B**). This indicates that an additional magnetic interaction is competing with the AFM

interaction. (**C**) and (**F**) The first derivative of the magnetic susceptibility shown in (**B**) and (**E**), respectively, showing that there might be another magnetic transition below the $T_N$. (**G**)-(**L**) Field- and crystal orientation-dependent magnetization measured at different temperatures. The suggested magnetic structures are shown in each figure. (**G**) Magnetization for $H//z$ at 2 K. An interlayer canted AFM coupling, which is very close to AFM configuration, can be considered at the initial step (0 T), as the field increases, a spin-flip transition happens at around 0.08 T. Then the magnetization follows a hysteresis curve, where spin-flip transitions connect FM phase and AFM phase. It is noteworthy that the FM phase can be maintained even at 0 T, indicating that the interlayer AFM exchange coupling is rather weak and compete with an additional interaction which favors FM phase. (**H**) Magnetization for $H//z$ at 5 K. The situation is similar to that in (**G**), however the additional interaction, which favors FM phase, becomes much weaker and the hysteresis curve becomes much smaller and in a plait shape. At 0 T, the FM states cannot be sustained, instead, canted AFM states are more stable. (**I**) Magnetization for $H//z$ at 10 K. As the additional interaction, which favors FM phase, becomes even weaker than that in (**H**), the magnetic hysteresis disappears, only spin-flip transitions connect FM phase at high fields and AFM phase at low fields. (**J**)-(**L**) Magnetization for $H//xy$ at different temperatures. A small magnetic hysteresis can be observed at 2 K (in **J**), indicating that the interlayer AFM coupling is canted at 0 T, in agreement with the magnetic configuration suggested in **G**.

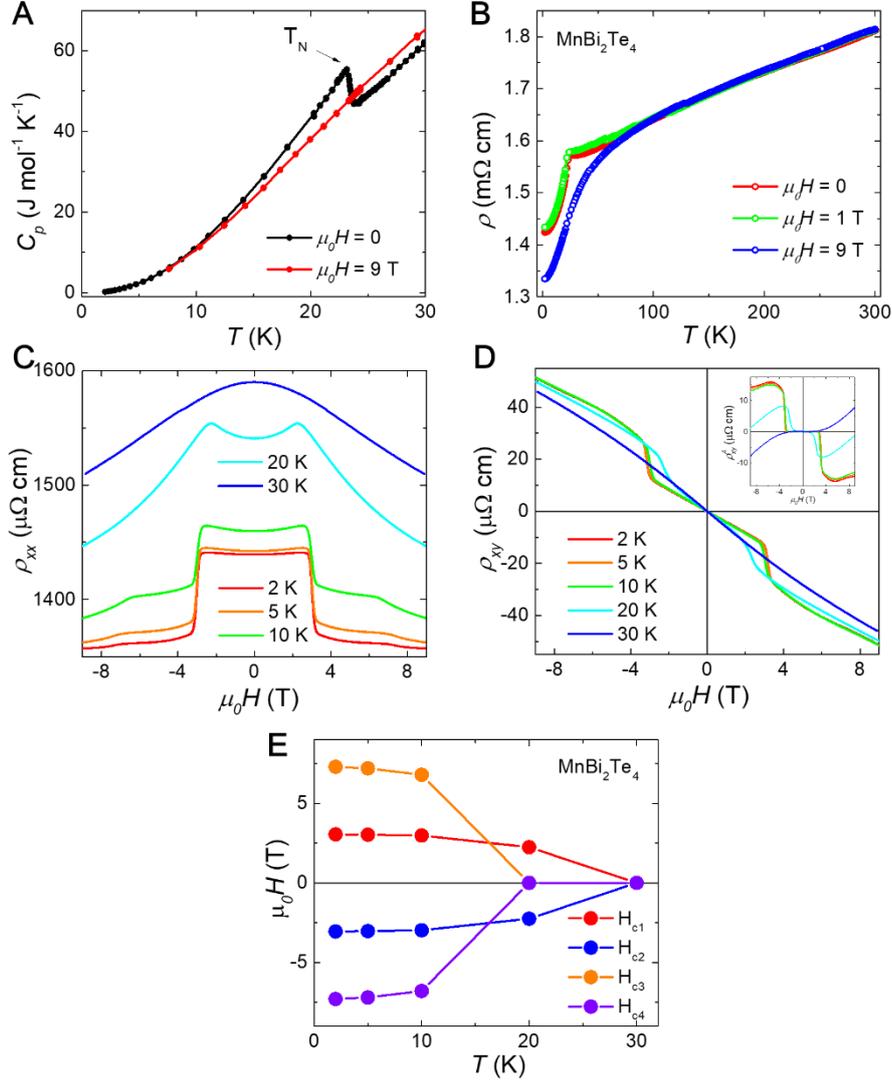

**Fig. S7. Additional physical properties of MnBi$_2$Te$_4$ single crystal.** (**A**) Low-temperature specific heat. (**B**) Resistivity. The AFM transition can be clearly seen by a kink in resistivity. (**C**) Magnetoresistance. (**D**) Hall resistivity. The inset shows the anomalous Hall resistivity. At low temperatures (below 10 K), there is a spin-flop transition (canted AFM transition) at ~3.4 T and a FM transition at ~7.3 T. For the electrical transport properties measurements, the current flowed in *a-b* plane, while the applied magnetic field was along *c* axis. (**E**) Temperature and field dependence of magnetic structures for MnBi$_2$Te$_4$ single crystal. Above 25 K ($T_N$), it is paramagnetic region. AFM region is between $H_{c1}$ and $H_{c2}$, canted AFM regions are between $H_{c1}$ and $H_{c3}$ for positive field and between $H_{c2}$ and $H_{c4}$ for negative field, and FM region is above $H_{c3}$ for positive field and below $H_{c4}$ for negative field.

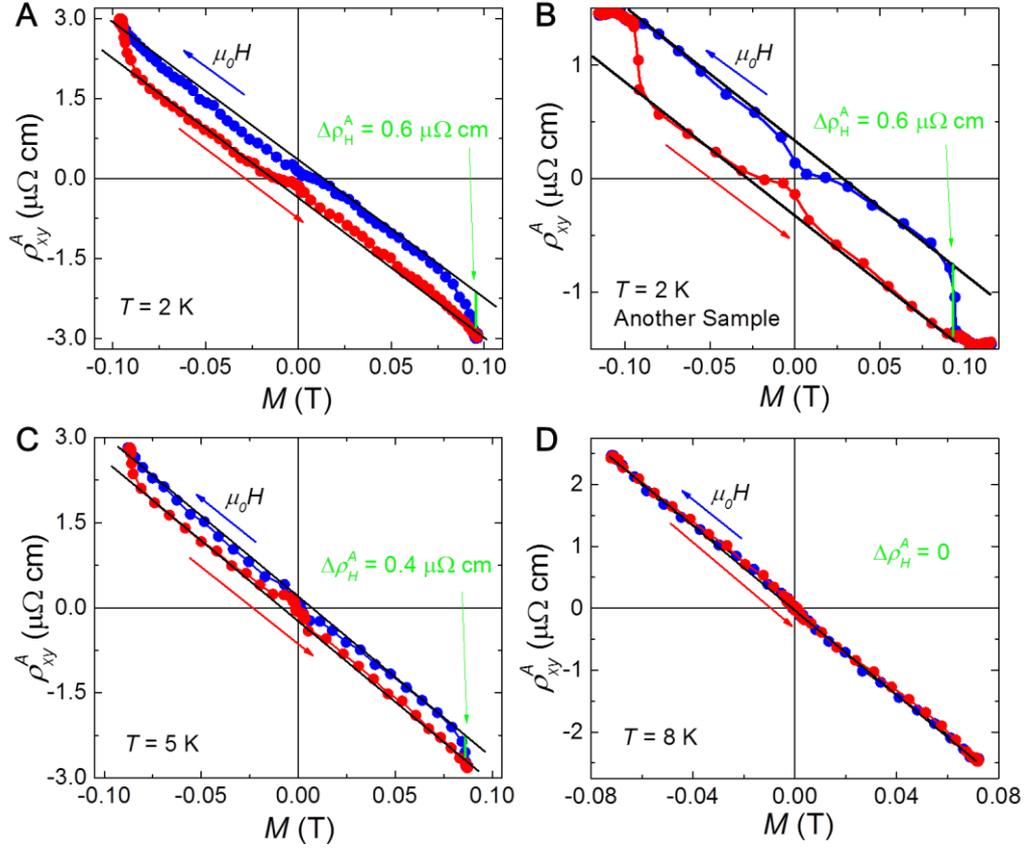

**Fig. S8. Magnetization dependence of $\rho_{yx}^A$ for MnBi$_4$Te$_7$ single crystal.** (**A**) 2 K. (**B**) 2 K for another sample with higher carrier concentration. (**C**) 5 K, (**D**) 8K. The anomalous Hall resistivity which is not coupled with magnetization ($\Delta\rho_H^A$) is probably due to the contribution from a net Berry curvature of a noncollinear spin texture (*37*).

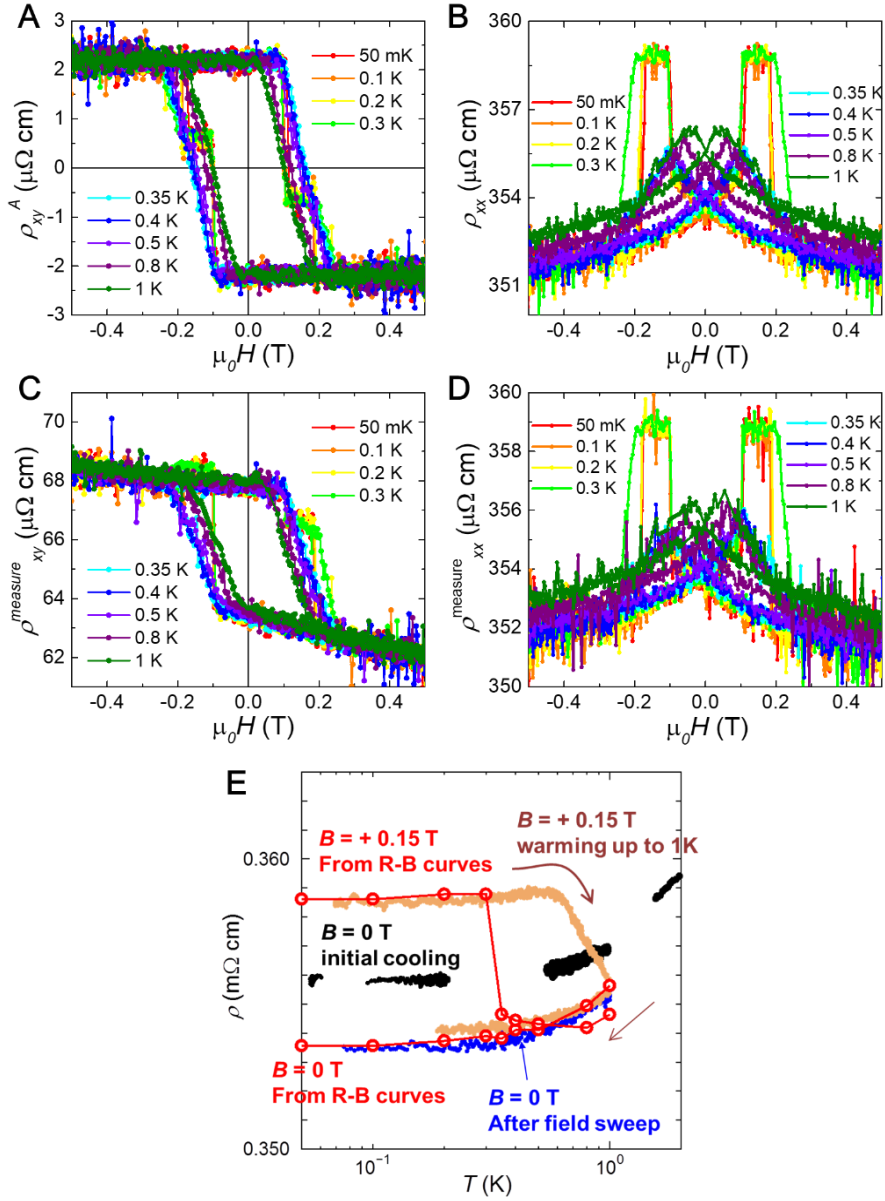

**Fig. S9. Additional electrical transport properties below 1K for MnBi$_4$Te$_7$ single crystal.** Field dependence of anomalous Hall resistivity (**A**) and longitudinal resistivity (**B**) below 1 K, obtained using a dilution refrigerator. (**C**) and (**D**) are the raw data of the measurements, where both of $\rho_{xy}^{\text{measure}}$ and $\rho_{xx}^{\text{measure}}$ contain the contributions from the Hall effect and magnetoresistance effect. This is mainly due to that the electrodes we made are not perfectly aligned. See Supplementary Text Section 1 for more information of the data analysis. (**E**) Temperature and field dependence of longitudinal resistivity. Four sets of data (different colors) will be interpreted. (1) The data shown in black were measured in the initial cooling at 0 T. (2) The data shown in blue were measured at 0 T after field sweep (from 1 T), showing a lower resistivity. This suggests that interlayer coupling is in a FM manner. (3) The data shown in brown were measured at 0.15 T (the field was increased from -1 T), following a temperature

increase from 70 mK to 1K and then a temperature decrease from 1K to 200 mK. When temperature increases from 70 mK to 600 mK, the sample shows a high-resistivity state, suggesting an AFM interlayer coupling. When temperature increases from 600 mK to 1 K, the resistivity drops quickly, crossing the resistivity in the initial cooling (data in black) and reaching the resistivity after field sweep (data in blue, low-resistivity state). When temperature decreases from 1 K to 200 mK, the sample shows a low-resistivity state, similar to the data in blue. Overall, the brown data indicate that, at 0.15 T, the AFM interlayer coupling is metastable and can be weakened by thermal energy, and a spin-flip transition (from 600 mK to 1 K) can take place when temperature increases. This result supports the assumption that the present materials system undergoes competing magnetic order. (4) The data shown by red circles were taken from **D**. The result is in agreement with the data shown in brown and blue. The discrepancy between the data in red and brown is possibly due to the heat release from the magnetic switching (switch 1: from FM phase, where magnetic moment is antiparallel to the applied field, to AFM phase), which may motivate the subsequent magnetic switching (switch 2: from AFM phase to FM phase, where magnetic moment is parallel to the applied field). Therefore, in field sweep (red), a smaller thermal energy (lower temperature) is required for switch 2, while in temperature sweep (brown) a higher temperature is need for switch 2. The presented data were smoothed by averaging every two data points of the raw data.

**Table S1. Crystal structure parameters of Mn-Bi-Te compounds.** Fractional atomic coordinates, site occupancies, and isotropic atomic displacement parameters $U_{eq}$ at room temperature for MnBi$_2$Te$_4$ (space group $R\bar{3}m$, a = b = 4.32474(9) Å, $c$ = 40.87999(3) Å), MnBi$_4$Te$_7$ (space group $P\bar{3}m1$, a = b = 4.35840(6) Å, $c$ = 23.78051(4) Å), and MnBi$_6$Te$_{10}$ (space group $R\bar{3}m$, a = b = 4.36582(5) Å, $c$ = 101.76660(9) Å) obtained from powder XRD analyses shown in Fig. S1.

**MnBi$_2$Te$_4$** (wRp = 0.1257 and Rp = 0.0835)

| Atom | Site | $x$ | $y$ | $z$ | Occupancy | $U_{eq}$ |
|---|---|---|---|---|---|---|
| Te1 | 6c | 0 | 0 | 0.1351(7) | 1 | 0.02 |
| Te2 | 6c | 0 | 0 | 0.2946(5) | 1 | 0.02 |
| Bi1 | 6c | 0 | 0 | 0.4250(3) | 0.913(3) | 0.02 |
| Mn1 | 6c | 0 | 0 | 0.4250(3) | 0.087(2) | 0.02 |
| Bi2 | 3a | 0 | 0 | 0 | 0.234(1) | 0.02 |
| Mn2 | 3a | 0 | 0 | 0 | 0.766(2) | 0.02 |

**MnBi$_4$Te$_7$** (wRp = 0.0833 and Rp = 0.0605)

| Atom | Site | $x$ | $y$ | $z$ | Occupancy | $U_{eq}$ |
|---|---|---|---|---|---|---|
| Te1 | 1b | 0 | 0 | 1/2 | 1 | 0.02 |
| Te2 | 2d | 1/3 | 2/3 | 0.0644(2) | 1 | 0.02 |
| Te3 | 2c | 0 | 0 | 0.2359(3) | 1 | 0.02 |
| Te4 | 2d | 1/3 | 2/3 | 0.3410(2) | 1 | 0.02 |
| Bi1 | 2d | 1/3 | 2/3 | 0.8407(4) | 1 | 0.02 |
| Bi2 | 4h | 1/3 | 2/3 | 0.5850(6) | 1 | 0.02 |
| Mn1 | 1a | 0 | 0 | 0 | 1 | 0.02 |

**MnBi$_6$Te$_{10}$** (wRp = 0.0833 and Rp = 0.0605)

| Atom | Site | $x$ | $y$ | $z$ | Occupancy | $U_{eq}$ |
|---|---|---|---|---|---|---|
| Te5 | 6c | 0 | 0 | 0.1802(2) | 1 | 0.02 |
| Te4 | 6c | 0 | 0 | 0.1165 (1) | 1 | 0.02 |
| Te3 | 6c | 0 | 0 | 0.4143(4) | 1 | 0.02 |
| Te2 | 6c | 0 | 0 | 0.0533(0) | 1 | 0.02 |
| Te1 | 6c | 0 | 0 | 0.3492(5) | 1 | 0.02 |
| Bi3 | 6c | 0 | 0 | 0.4698 (1) | 1 | 0.02 |
| Bi2 | 6c | 0 | 0 | 0.2354(1) | 1 | 0.02 |
| Bi1 | 6c | 0 | 0 | 0.2971(6) | 1 | 0.02 |
| Mn1 | 3a | 0 | 0 | 0 | 1 | 0.02 |